\documentclass[prb,aps,epsf,twocolumn,superscriptaddress,showpacs,10pt]{revtex4-2}
\usepackage{graphicx,amsfonts,times,bm,amsmath,verbatim,color,array}
\usepackage{xcolor}
\usepackage{algorithmic}
\usepackage[colorlinks,urlcolor=blue,citecolor=blue,linkcolor=blue]{hyperref}
\usepackage[capitalize]{cleveref}
\usepackage{braket}
      
\usepackage{multirow}
\usepackage[all]{hypcap} 
\usepackage{xspace}
\usepackage{amssymb}
\usepackage{amsmath}
\usepackage{appendix}
\usepackage{pifont}
\usepackage{booktabs}
\usepackage{etoolbox} 

\AtBeginDocument{%
  \heavyrulewidth=.08em
  \lightrulewidth=.05em
  \cmidrulewidth=.03em
  \belowrulesep=.65ex
  \belowbottomsep=0pt
  \aboverulesep=.4ex
  \abovetopsep=0pt
  \cmidrulesep=\doublerulesep
  \cmidrulekern=.5em
  \defaultaddspace=.5em
}

\begin{document}

\title{Quantum Geometry Driven Optical Responses in 1T-$\mathrm{MX}_2$ Monolayers: A Symmetry-Constrained Slater–Koster Tight-Binding Approach }

\author{Bikram Baruah}
\email{bikram22$_$rs@phy.nits.ac.in}
\affiliation{Department of Physics, National Institute of Technology Silchar, Assam 788010, India}

\author{Snehasish Nandy}
\email{snehasish@phy.nits.ac.in}
\affiliation{Department of Physics, National Institute of Technology Silchar, Assam 788010, India}

\author{Subhasis Panda}
\email{subhasis@phy.nits.ac.in}
\affiliation{Department of Physics, National Institute of Technology Silchar, Assam 788010, India}

\begin{abstract}
Centrosymmetric 1T-$\mathrm{MX}_2$ monolayers have attracted considerable attention owing to their intriguing transport properties and potential technological applications arising from the interplay among quantum geometry, electronic band structure, and band-gap characteristics. Despite this rich physics, a comprehensive microscopic tight-binding description that simultaneously captures these properties remains insufficiently established, while first-principles approaches can be computationally demanding for systematic investigations across different materials and perturbations. Here, we develop a transferable eleven-band Slater--Koster tight-binding description of monolayer 1T-$\mathrm{MX}_2$ transition-metal dichalcogenides and use it to establish a connection among their microscopic electronic structure, quantum geometry, and optical response. The model is constructed in an orthogonal orbital basis from the crystal geometry and symmetry-constrained Slater--Koster parameters, with material-specific parametrizations obtained from first-principles calculations for monolayer $\mathrm{ZrS}_2$ and $\mathrm{HfS}_2$. The resulting geometry-based Hamiltonian accurately describes the low-energy electronic structures and provides a natural framework for extending the analysis to the broader isostructural 1T-$\mathrm{MX}_2$ family. We find that the pristine monolayers possess a finite quantum metric, with the dominant contribution concentrated in the two highest occupied bands owing to the small near-gap energy separation and strong metal--chalcogen $p$--$d$ hybridization. Furthermore, we verify the interband $f$-sum rule, which directly relates the integrated optical spectral weight to the Brillouin-zone-averaged quantum metric. Our results establish optical spectral weight as an experimentally accessible probe of the quantum geometry of occupied Bloch states and provide a unified microscopic framework for connecting electronic structure, quantum geometry, and measurable optical responses across the 1T-$\mathrm{MX}_2$ family.
\end{abstract}

\maketitle

%=============================================================
\section{Introduction}
\label{sec:intro}
%=============================================================
Layered transition-metal dichalcogenides (TMDs) in the octahedral 1T phase have attracted increasing attention as a distinct class of two-dimensional semiconductors with electronic properties that differ markedly from those of their more extensively studied 2H counterparts. Among them, monolayer $\mathrm{ZrS}_2$ and $\mathrm{HfS}_2$ exhibit indirect band gaps in the technologically relevant 1--1.5 eV range, along with relatively low carrier effective masses and good environmental stability, making them promising candidates for optoelectronic and photovoltaic applications~\cite{PRB.100.165304,PRM.3.074001,Wei2023,Mattinen2019}. Beyond their intrinsic properties, the band gap is highly sensitive to layer thickness and strain~\cite{Orujlu2026,Zhang2022}, offering a practical route to tailor the optical response, and both the compounds have already been demonstrated as active layers in high-performance photodetectors and phototransistors~\cite{Mattinen2019,Wang2018,Tanthirige2019}. Their electronic structures exhibit pronounced orbital selectivity, with valence states derived predominantly from chalcogen $p$ orbitals and conduction states dominated by transition-metal $d$ orbitals, together with substantial $p$--$d$ hybridization. This orbital selectivity gives rise to characteristic interband optical features and makes these materials useful platforms for investigating the interplay among electronic structure, optical response, and lattice geometry.

Beyond these conventional electronic and optical properties, the quantum geometry of Bloch states has emerged as an essential ingredient in understanding optical and transport phenomena in condensed matter physics~\cite{Liu2025,Raquel2026,Gao2025}. While the Berry curvature has long been recognized as a central quantity in this context, the quantum metric, corresponding to the real part of the quantum geometric tensor, has emerged as an important complementary descriptor of Bloch-state geometry, with implications ranging from flat-band superconductivity and nonlinear optical response to electron--phonon coupling~\cite{Yu2025,Yu2024,Torma2023essay}. This theoretical framework has recently gained direct experimental support through the first measurement of quantum metric tensor in a solid, obtained by photoemission spectroscopy on black phosphorus~\cite{Kim2025}, while nonlinear Hall measurements in a topological antiferromagnetic heterostructure have revealed its contribution to transport even when the Berry curvature is forbidden by symmetry~\cite{Gao2023}. On the theoretical side, analytic treatments that link the quantum metric to the optical conductivity~\cite{Ezawa2024} and generalized optical sum rules~\cite{Nishchhal2024,Verma2025} have established its connection to measurable spectral weight. In pristine 1T-$\mathrm{MX}_2$ monolayers, the coexistence of inversion and time-reversal symmetry forces the Berry curvature to vanish throughout the Brillouin zone, leaving the quantum metric as the sole non-vanishing contribution to the quantum geometric tensor. These systems therefore provide a particularly clean setting for isolating the quantum-metric contribution to the optical response and establishing its connection to measurable optical spectral weight.

A quantitative yet computationally efficient description of the low-energy electronic structure is therefore desirable for treating the electronic, optical, and quantum-geometric properties within a common framework, while retaining explicit control over the underlying orbital and hopping parameters. Tight-binding (TB) models provide such a framework by expressing the electronic structure in terms of localized orbitals and parametrized hopping amplitudes fitted to first-principles calculations~\cite{Marzari2012,Papaconstantopoulos2003,Jorissen2024}. In particular, Slater--Koster (SK) formulations provide a physically transparent description of this approach, in which the two-center hopping integrals are expressed as interpretable functions of the bond geometry, directly connecting the Hamiltonian to the underlying lattice structure~\cite{Slater-Koster}. Early semi-empirical SK descriptions of the 1T-$\mathrm{MX}_2$ family were developed by Murray \textit{et al.}, who employed a non-orthogonal basis fitted to experimental optical data~\cite{Murray1972}. Although non-orthogonal bases offer additional flexibility, an orthogonal basis can retain the essential low-energy physics while avoiding the additional complexity associated with the overlap matrix. Orthogonal SK models parametrized directly from DFT have consequently been developed for several 2H-TMD compounds, demonstrating the importance of longer-range hopping processes and providing efficient platforms for investigating strain and transport effects~\cite{Cappelluti2013,Roldan2014,Ridolfi2015,Dias2018,Silva-Guillen2016,Peng2024}. In parallel, Wannier-based Hamiltonians have been successfully applied to 2H TMDs, providing an accurate first-principles description of nanostructures, transport, and superlattices~\cite{Lado_2016,Ghosh2025}. Despite these advances, a unified orthogonal SK framework for the 1T-$\mathrm{MX}_2$ family that simultaneously provides a quantitatively DFT-constrained electronic structure and a direct route to optical and quantum-geometric observables remains lacking.

In this work, we develop a generalized eleven-band SK TB framework for monolayer 1T-$\mathrm{MX}_2$ compounds, based on an orthogonal basis comprising the five transition-metal $d$ orbitals and six chalcogen $p$ orbitals. The Hamiltonian is constructed from the lattice geometry and a symmetry-constrained set of SK hopping integrals, providing a compact framework suitable for extension across the 1T-$\mathrm{MX}_2$ family. We demonstrate the approach for monolayer $\mathrm{ZrS}_2$ and $\mathrm{HfS}_2$ by fitting the low-energy electronic structure to first-principles calculations. The resulting model accurately describes the band gaps, orbital character, and crystal-field splittings of the DFT electronic structure while providing the computational efficiency needed for extensions to large-scale systems, including strain-engineered heterostructures and moir\'e superlattices involving 1T-$\mathrm{MX}_2$ materials. Building on this Hamiltonian, we calculate the optical conductivity and quantum metric of the occupied bands and establish their connection through the interband $f$-sum rule, thereby linking the geometry of the occupied Bloch states to measurable optical spectral weight. This unified description establishes a direct link between the microscopic electronic structure and the measurable optical response through the quantum geometry of the occupied Bloch states. 

The paper is organized as follows. Sections~\ref{sec:theory} and~\ref{sec:model} describe the first-principles calculations, construction of the eleven-band TB Hamiltonian, and SK parametrization. Section~\ref{sec:electronic_optical} presents the electronic and optical properties obtained from the model as well as the quantum geometric analysis and its connection to optical spectral weight through the interband $f$-sum rule. Finally, Sec.~\ref{sec:conclusion} summarizes the main findings and outlines possible future directions.

%=============================================================
\section{Crystal Structure and Tight-Binding Model}
\label{sec:theory}
%=============================================================
%-------------------------------------------------------------
\subsection{Crystal Structure and Lattice Geometry}
\label{subsec:crystal}
\begin{figure}[h]
\centering
\includegraphics[width=\columnwidth]{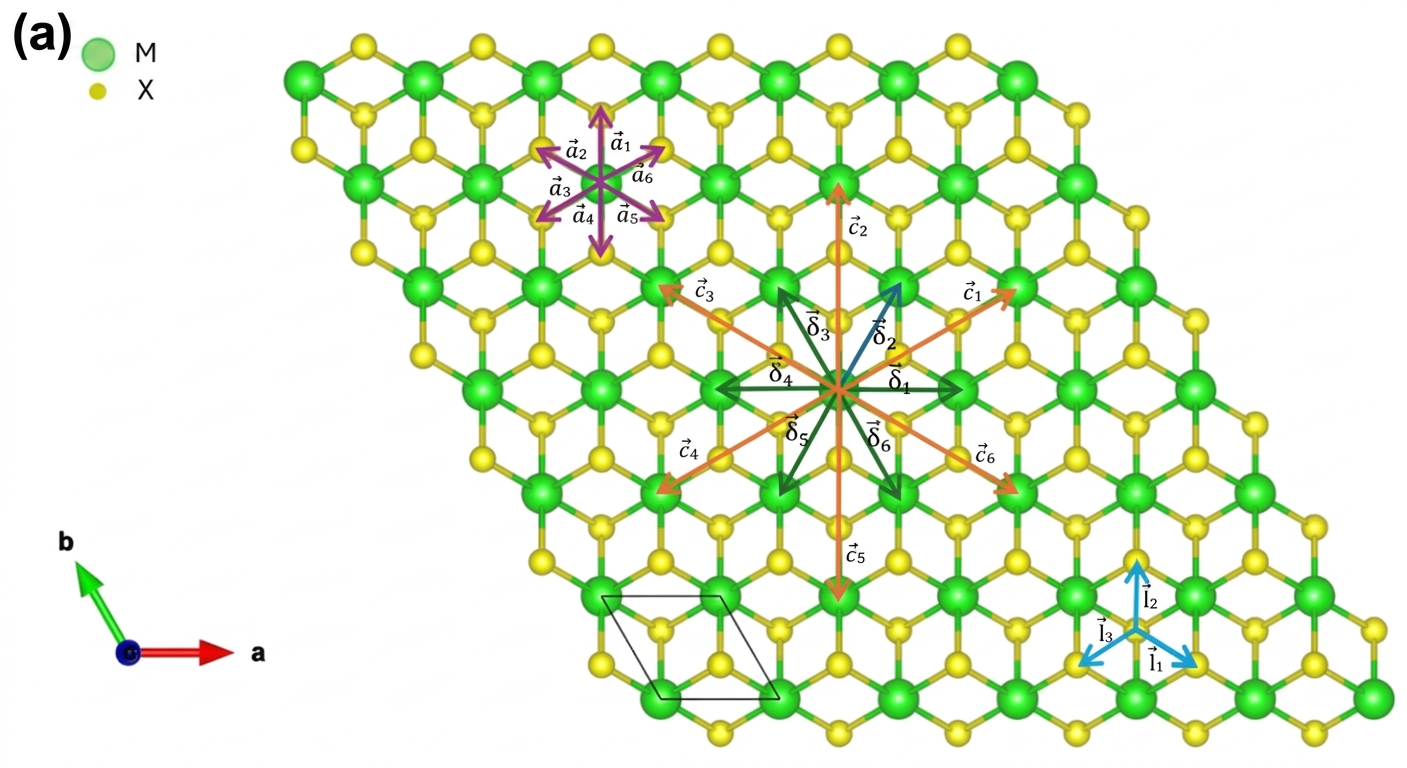}
\includegraphics[width=\columnwidth]{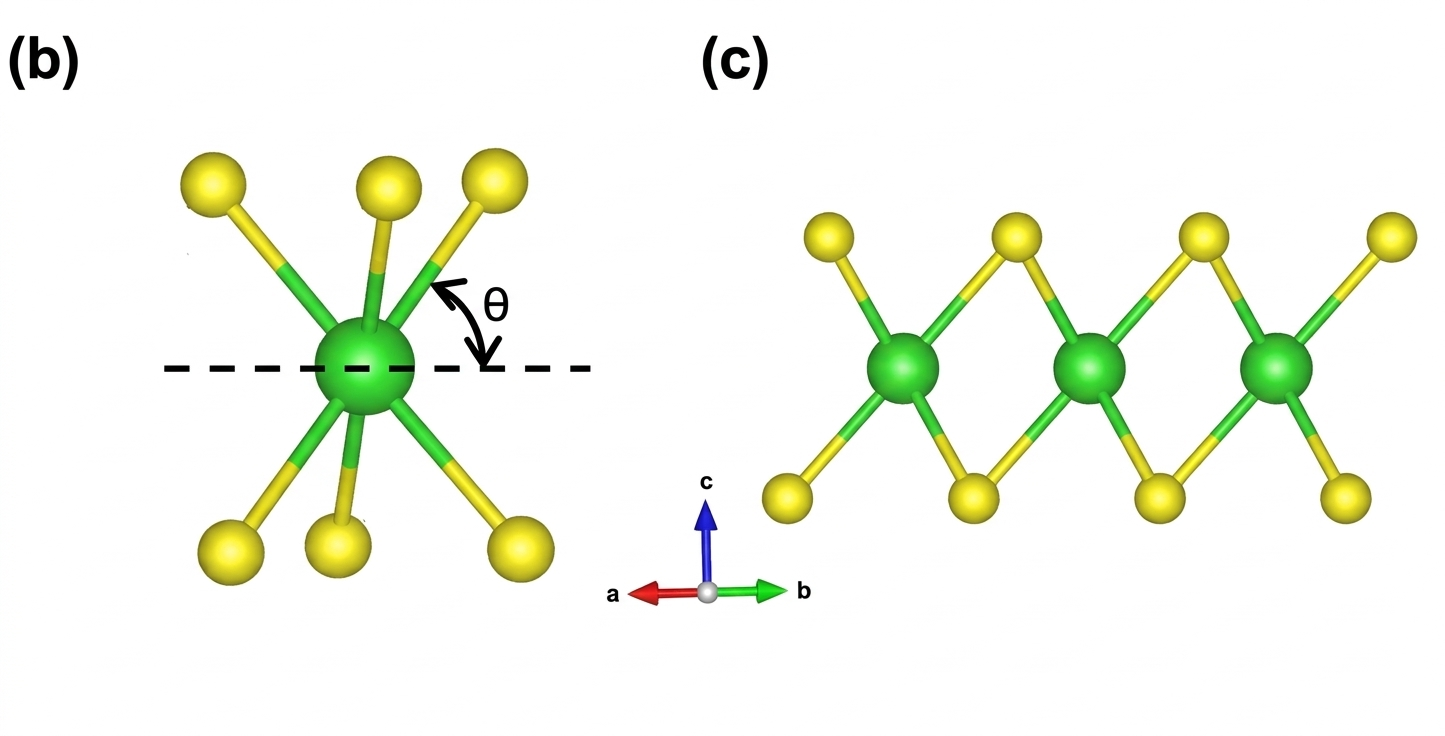}
\caption{(a) Crystal structure of a monolayer 1T-$\mathrm{MX}_2$ compound, showing hopping vectors connecting nearest and next-nearest neighbor sites in the basal plane, (b) the angle between M--X bond and M layer, and (c) side view of the single layer 1T-$\mathrm{MX}_2$ compounds showing X--M--X stacking}
\label{fig:structure}
\end{figure}

%-------------------------------------------------------------
All 1T-$\mathrm{MX}_2$ compounds considered in this study share the same basic structure, consisting of a single layer with three atomic planes arranged in an X–M–X sequence, where the metal (M) layer is sandwiched between two chalcogen (X) layers. Each metal atom is surrounded by six chalcogen neighbors, with three above and three below, forming a \emph{trigonal antiprismatic} coordination. This arrangement, illustrated schematically in Fig.~\ref{fig:structure}, can be viewed as a trigonal distortion of an ideal octahedron ($O_h$). The two triangular chalcogen layers are rotated by $60^\circ$ relative to each other, which effectively compresses the octahedral symmetry along the threefold $C_3$ axis and lowers the point-group symmetry to $D_{3d}$, corresponding to the space group $P\bar{3}m1$ (No.~164).

The reduction from $O_h$ to $D_{3d}$ symmetry has direct consequences for the electronic structure. In $O_h$, the five $d$ orbitals split into $t_{2g}$ and $e_g$. Under $D_{3d}$, the $t_{2g}$ manifold further splits into an $a_{1g}$ singlet and an $e_g^{(1)}$ doublet, while the $e_g$ doublet becomes $e_g^{(2)}$;
$O_h: t_{2g} \oplus e_g \longrightarrow
D_{3d}: a_{1g} \oplus e_g^{(1)} \oplus e_g^{(2)}.
$
In 1T-$\mathrm{MX}_2$, this places the $e_g^{(1)}$ doublet at the bottom of the conduction band at $M$ point, followed by the $a_{1g}$ singlet, consistent with both DFT and TB results. Here, $a_{1g}$ corresponds to $d_{z^2}$, $e_g^{(1)}$ to $\{d_{xz}, d_{yz}\}$, and $e_g^{(2)}$ to $\{d_{x^2-y^2}, d_{xy}\}$. The trigonal antiprismatic coordination is the key structural feature that distinguishes the 1T phase of $\mathrm{MX}_2$ compounds from that of the trigonal prismatic ($D_{3h}$) structure of the 2H phase. 

\begin{table}[h]
\caption{Structural parameters of the 1T-$\mathrm{MX}_2$ compounds studied in this work: the in-plane lattice constant $a$ and the angle $\theta$ between the M--X bond and the M plane.}
\label{tab:structure}
\begin{tabular}{lcc}
\toprule
Compound \quad & $a$ (\AA) \quad & $\theta$ (rad) \\
\midrule
$\mathrm{ZrS}_2$  & 3.69 \quad & 0.5965 \\
$\mathrm{HfS}_2$  & 3.652 \quad & 0.5993 \\
\bottomrule
\end{tabular}
\end{table}

In the basal plane, the atoms form a triangular Bravais lattice with lattice constant $a$ (see Table~\ref{tab:structure}), with primitive vectors $\vec{R}_1 \!=\! a\hat{x}$ and $\vec{R}_2 \!=\! -\tfrac{a}{2}\hat{x} + \tfrac{a\sqrt{3}}{2}\hat{y}$. The corresponding reciprocal vectors are $\vec{b}_1 \!=\! \frac{2\pi}{a}(\hat{k}_x + \tfrac{1}{\sqrt{3}}\hat{k}_y)$ and $\vec{b}_2 \!=\! \tfrac{2\pi}{a}(\tfrac{2}{\sqrt{3}}\hat{k}_y)$, defining a hexagonal Brillouin zone with high-symmetry points $\Gamma \!=\! (0,0)$, $M \!=\! (\tfrac{\pi}{a}, \tfrac{\pi}{\sqrt{3}a})$, and $K \!=\! (\tfrac{4\pi}{3a}, 0)$.

The out-of-plane geometry is fully determined by the M–X bond angle $\theta$, defined with respect to the metal plane (Fig.~\ref{fig:structure}(b)). For an ideal octahedron, $\cos\theta \!=\! \sqrt{2/3}$, giving $\theta \approx 0.6154$ radian. In real compounds, $\theta$ deviates slightly due to differences in ionic radii, with values listed in Table~\ref{tab:structure}. This angle sets the bond direction cosines and thus governs the magnitude and spatial anisotropy of the orbital overlaps. The M–X bond length is $b = a/(\sqrt{3}\cos\theta)$, and the corresponding hopping vectors are given in Sec.~\ref{subsec:vectors}.

In the $P\bar{3}m1$ structure (No.~164), the metal atom occupies the $1a$ site and the two chalcogen sublattices occupy the $2d$ sites. Together with the relaxed bond angle $\theta$, these crystallographic constraints uniquely determine the orbital embedding used in the SK Hamiltonian, thereby eliminating ambiguity in the resulting geometry-dependent optical and quantum-geometric observables~\cite{Telle2026,Simon2020}.

%-------------------------------------------------------------
\subsection{DFT Calculations}
\label{subsec:dft}
%-------------------------------------------------------------

All first-principles calculations were performed within density functional theory (DFT) using the Vienna Ab initio Simulation Package (VASP)~\cite{vasp93, vasp96}. The Perdew-Burke-Ernzerhof (PBE) generalized gradient approximation was used for the exchange-correlation functional~\cite{PBE}. The Kohn-Sham equations were solved using a plane-wave basis set with an energy cutoff of 500~eV. Atomic positions and cell parameters were relaxed until the residual Hellmann-Feynman forces on every atom fell below $5\times10^{-4}$~eV/\AA\, and the total-energy convergence criterion was set to $10^{-8}$~eV. The Brillouin zone was sampled with an $18\times18\times1$ Monkhorst-Pack $k$-grid during structural relaxation. To generate the 2D monolayer characteristic, a vacuum layer of at least 30~\AA\ along the $z$ direction was included in all calculations.

\begin{figure}[h]
\centering
\includegraphics[width=\columnwidth]{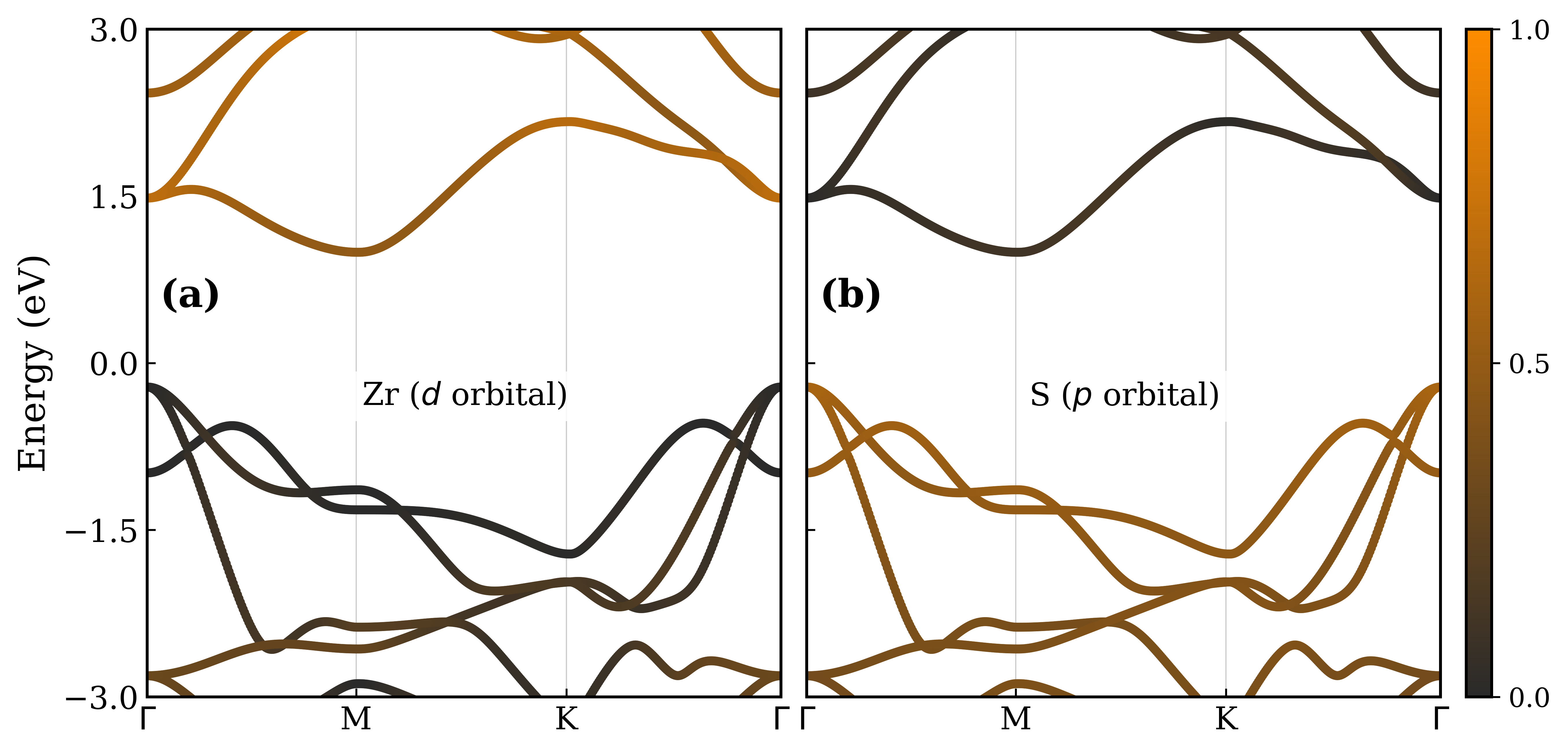}
\caption{DFT-computed orbital-projected band structures of monolayer $\mathrm{ZrS}_2$, illustrating the partial contributions of (a) Zr 4$d$ and (b) S 3$p$ states.}
\label{fatband ZrS2}
\label{fig:dft_bands}
\end{figure}

The orbital-projected band structure for monolayer $\mathrm{ZrS}_2$ was calculated along the $\Gamma$--M--K--$\Gamma$ path and is shown in Fig. \ref{fatband ZrS2}. Inspection of the orbital weights confirms that the valence and conduction bands in the energy window $\pm 4$~eV around the Fermi level are dominated by the five Zr $4d$ orbitals ($d_{z^2}$, $d_{x^2-y^2}$, $d_{xy}$, $d_{yz}$, $d_{zx}$) and the three S $3p$ orbitals ($p_x$, $p_y$, $p_z$) of each chalcogen atom. This establishes the appropriate Hilbert space for our TB model, as described in Sec.\ref{subsec:crystal}.
All 1T compounds considered in this study are \emph{indirect} band gap semiconductors, with the valence band maximum (VBM) and the conduction band minimum (CBM) located at high symmetry k-points $\Gamma$ and M, respectively.

\begin{table}[h]
\caption{Comparison of DFT and TB band gaps for monolayer 1T-$\mathrm{MX}_2$ compounds. All materials exhibit indirect band gaps.}
\label{tab:gaps}
\begin{tabular}{lccc}
\toprule
 & \multicolumn{2}{c}{DFT(PBE) gap (eV)} &  \\
\cline{2-3}
Compound & This work & Literature & TB gap (eV) \\
\midrule
$\mathrm{ZrS}_2$  & 1.21 & 1.19~\cite{PRB.100.165304}, 1.21~\cite{PRM.3.074001} & 1.20 \\
$\mathrm{HfS}_2$  & 1.34 & 1.31~\cite{PRB.100.165304}, 1.36~\cite{PRM.3.074001} & 1.32 \\
\bottomrule
\end{tabular}
\end{table}

%-------------------------------------------------------------
\subsection{Hopping Vectors}
\label{subsec:vectors}
%-------------------------------------------------------------
Constructing the TB Hamiltonian requires specification of all relevant hopping pathways connecting neighboring atomic sites. We categorize these into three types: metal-chalcogen (M--X), metal-metal (M--M), and chalcogen-chalcogen (X-X) interactions.

Each metal (M) atom bonds with six chalcogen (X) neighbors, with three located in the upper ($+z$) sublattice and three in the lower ($-z$) sublattice, consistent with octahedral coordination. With the M atom placed at the origin, the six M--X nearest-neighbor (NN) vectors $\vec{a}_i$($i=1\dots6$) are expressed as:
\begin{align}
\begin{gathered}
\vec{a}_1 = b(0,\, \cos\theta,\, -\sin\theta), \hfill\\
\vec{a}_2 = b\!\left(-\tfrac{\sqrt{3}}{2}\cos\theta,\,
                       \tfrac{1}{2}\cos\theta,\, \sin\theta\right),\hfill\\
\vec{a}_3 = b\!\left(-\tfrac{\sqrt{3}}{2}\cos\theta,\,
                       -\tfrac{1}{2}\cos\theta,\, -\sin\theta\right),\hfill\\
\vec{a}_4 = b(0,\, -\cos\theta,\, \sin\theta), \hfill\\
\vec{a}_5 = b\!\left(\tfrac{\sqrt{3}}{2}\cos\theta,\,
                       -\tfrac{1}{2}\cos\theta,\, -\sin\theta\right),\hfill\\
\vec{a}_6 = b\!\left(\tfrac{\sqrt{3}}{2}\cos\theta,\,
                       \tfrac{1}{2}\cos\theta,\, \sin\theta\right),\hfill
\label{eq:a_vectors}
\end{gathered}
\end{align}
where $b = a/(\sqrt{3}\cos\theta)$ is the M--X bond length. The vectors $\vec{a}_1$, $\vec{a}_3$, and $\vec{a}_5$ connect the M atom to the lower ($-z$) chalcogen sublattice, while $\vec{a}_2$, $\vec{a}_4$, and $\vec{a}_6$ connect it to the upper ($+z$) sublattice. Under lattice inversion $\mathbf{r} \to -\mathbf{r}$, the 1T inversion center relates upper and lower M--X bond vectors pairwise: $\vec{a}_i = -\vec{a}_{i+3}$, for $i=1,2,3$.

In the case of M--M and X--X hoppings, all bond vectors are confined to the basal plane. The six nearest-neighbor (NN) vectors $\vec{\delta}_i$ link atoms within the same triangular sublattice, separated by a distance $a$:
\begin{align}
\begin{gathered}
\vec{\delta}_1 = a(1,\, 0,\, 0), \hfill\\
\vec{\delta}_2 = a\!\left(\tfrac{1}{2},\, \tfrac{\sqrt{3}}{2},\, 0\right),
\hfill\\
\vec{\delta}_3 = a\!\left(-\tfrac{1}{2},\, \tfrac{\sqrt{3}}{2},\, 0\right), \hfill
\label{eq:delta_vectors}
\end{gathered}
\end{align}
with $\vec{\delta}_{i+3} = -\vec{\delta}_i$ for $i = 1, 2, 3$.
The six next-nearest-neighbor (NNN) vectors $\vec{c}_i$ connect atoms separated by a distance $d = a\sqrt{3}$:
\begin{align}
\begin{gathered}
\vec{c}_1 = d\!\left(\tfrac{\sqrt{3}}{2},\, \tfrac{1}{2},\, 0\right), \hfill\\
\vec{c}_2 = d(0,\, 1,\, 0), \hfill\\
\vec{c}_3 = d\!\left(-\tfrac{\sqrt{3}}{2},\, \tfrac{1}{2},\, 0\right),\hfill
\label{eq:c_vectors}
\end{gathered}
\end{align}
with $\vec{c}_{i+3} = -\vec{c}_i$.

The final set of bond vectors connects the two chalcogen sublattices directly. The three nearest-neighbor inter-sublayer X$_t$–X$_b$ vectors are: 
\begin{align}
\begin{gathered}
\vec{l}_1 = b\!\left(\tfrac{\sqrt{3}}{2} \cos\theta,\; \tfrac{1}{2} \cos\theta,\; -2\sin\theta \right), \hfill\\
\vec{l}_2 = b\!\left(0,\; -\cos\theta,\; -2\sin\theta \right), \hfill\\
\vec{l}_3 = b\!\left(-\tfrac{\sqrt{3}}{2} \cos\theta,\; \tfrac{1}{2} \cos\theta,\; -2\sin\theta \right),\hfill
\label{eq:V_vecs}
\end{gathered}
\end{align}
which connect the top and bottom sublattices and define the hopping block $\mathcal{H}_{tb}$. The magnitude of $\vec{l}_i$ is $|\vec{l}_i| = b\sqrt{\cos^2\theta + 4\sin^2\theta}$, which becomes $b\sqrt{2}$ in the ideal octahedral limit where $\cos\theta=\sqrt{2/3}$. Together with the material parameters listed in Table~\ref{tab:structure}, the lattice geometry defined in Eqs.~(\ref{eq:a_vectors}-\ref{eq:V_vecs}) provides the foundation for constructing the momentum-space Hamiltonian through the Bloch's theorem and the Fourier transformation.

%-------------------------------------------------------------
\subsection{Tight-Binding Hamiltonian}
\label{subsec:model}
%-------------------------------------------------------------
The primitive unit cell of a 1T-$\mathrm{MX}_2$ monolayer contains one M atom and two X atoms (one per sublattice). Based on the DFT orbital analysis presented in Sec.\ref{subsec:dft}, we include the five M $4d$ orbitals and six X $3p$ orbitals (three per chalcogen sublattice), yielding an eleven-dimensional Hilbert space per $\mathbf{k}$-point. Explicitly labeling the top ($t$) and bottom ($b$) sublattices, the basis is given by:
\begin{equation*}
\phi^\dagger =
\bigl(p^t_x,\, p^t_y,\, p^t_z,\,
      d_{z^2},\, d_{x^2-y^2},\, d_{xy},\, d_{yz},\, d_{zx},\,
      p^b_x,\, p^b_y,\, p^b_z\bigr).
\label{eq:raw_basis}
\end{equation*}
We adopt the shorthand notation $d_0 \equiv d_{z^2}$, $d_1 \equiv \{d_{yz}, d_{zx}\}$, and $d_2 \equiv \{d_{x^2-y^2}, d_{xy}\}$ to reflect the crystal-field splitting of the $d$ manifold in the octahedral ($O_h$) environment, which is reduced to $D_{3d}$ by the trigonal distortion.  The model is implemented and solved directly in the basis defined by the expression above.

In this basis, the TB Hamiltonian can be written in real space as~\cite{Peng2024}:
\begin{equation}
H=\sum_{i,\mu\nu}\Delta_{\mu\nu}c^\dagger_{i\mu}c_{i\nu}
  +\sum_{ij,\mu\nu}\!\bigl[t_{ij,\mu\nu}\,c^\dagger_{i\mu}c_{j\nu}+\mathrm{H.c.}\bigr],
\label{eq8}
\end{equation}
where $c^\dagger_{i\mu}$ ($c_{i\nu}$) is the creation (annihilation) operator for orbital $\mu$ ($\nu$) on site $i$, $\Delta_{\mu\nu}$ are the on-site energies, and $t_{ij,\mu\nu}$ are the hopping integrals. After the Fourier transformation, the momentum-space Hamiltonian takes the $11\times11$ block form
\begin{equation}
\mathcal{H}_{TB} \!=\!
\begin{pmatrix}
\Delta_{X_t} + \mathcal{H}_{tt} \!&\!
    \mathcal{H}^\dagger_{Mt}\!&\!
    \mathcal{H}_{tb} \\[4pt]
\mathcal{H}_{Mt} \!&\!
    \Delta_M + \mathcal{H}_{MM} \!&\!
    \mathcal{H}_{Mb} \\[4pt]
\mathcal{H}^\dagger_{tb} \!&\!
    \mathcal{H}^\dagger_{Mb} \!&\!
    \Delta_{X_b} + \mathcal{H}_{bb}
\end{pmatrix},
\label{eq:H_block}
\end{equation}
where the rows and columns are partitioned into the top chalcogen, metal, and bottom chalcogen subspaces, $(X_t|M|X_b)$. For brevity, the explicit dependence on $\mathbf{k}$ has been removed from the relevant terms. Here, $\mathcal{H}_{MM}$ is a $5\times5$ M–M hopping block, $\mathcal{H}_{tt}$ and $\mathcal{H}_{bb}$ are $3\times3$ intra-sublayer X–X hopping blocks, $\mathcal{H}_{Mt}$ and $\mathcal{H}_{Mb}$ are $5\times3$ metal–chalcogen coupling blocks, and $\mathcal{H}_{tb}$ is a $3\times3$ inter-sublayer X–X hopping block. The matrices $\Delta$ encode the on-site energies and are defined as:
\begin{align}
\begin{gathered}
\Delta_M = \text{diag} (E_{d_0}, E_{d_1}, E_{d_1}, E_{d_2}, E_{d_2}), \hfill
 \\
\Delta_{X_t} = \Delta_{X_b} = \text{diag} (E_{p_1}, E_{p_1}, E_{p_2}), \hfill
\end{gathered}
\end{align}
where $E_{d_l}$ $(l=0,1,2)$ are the crystal-field on-site energies of the three $d$-orbital groups, and $E_{p_1}$, $E_{p_2}$ are the on-site energies of the in-plane ($p_x,p_y$) and out-of-plane ($p_z$) chalcogen orbitals, respectively. The top and bottom sublattices share identical on-site energies ($\Delta_{X_t}=\Delta_{X_b}$), reducing the independent on-site parameters to five: $E_{d_0}$, $E_{d_1}$, $E_{d_2}$, $E_{p_1}$, $E_{p_2}$.

The individual hopping blocks in Eq.~\ref{eq:H_block} are assembled by Fourier-transforming the real-space hopping matrices as follows.

\textbf{In-plane M–M and intra-sublayer X–X hoppings.} Since the NN ($\vec{\delta}_i$) and NNN ($\vec{c}_i$) vectors occur in inversion symmetric pairs, $\vec{\delta}_{i+3}=-\vec{\delta}_i$ and $\vec{c}_{i+3}=-\vec{c}_i$, the corresponding Fourier sums reduce to cosine terms as:
\begin{align}
\mathcal{H}_{\upsilon} =
    2\sum_{i=1}^{3} \left [\cos(\mathbf{k}\cdot\vec{\delta}_i)\, T_\upsilon(\vec{\delta}_i)  + \cos(\mathbf{k}\cdot\vec{c}_i)\, T_\upsilon(\vec{c}_i) \right], 
\label{eq:H_inplane}
\end{align}
where $\upsilon \in \{MM, tt/bb\}$, with $T_{MM} \equiv T_M$ and $T_{tt/bb} \equiv T_X$. Here, $T_M(\vec{\delta}_i)$ and $T_M(\vec{c}_i)$ denote the NN and NNN M–M hopping matrices, while $T_X(\vec{\delta}_i)$ and $T_X(\vec{c}_i)$ are the corresponding intra-sublayer X–X hopping matrices.

\textbf{M–X coupling.} Since M and X sit on different sublattices, the three lower ($\vec{a}_1,\vec{a}_3,\vec{a}_5$) and three upper ($\vec{a}_2,\vec{a}_4,\vec{a}_6$) M–X bond vectors do not form inversion-symmetric pairs within a single hopping block.  The two M–X blocks are therefore built as independent complex-exponential sums:
\begin{align}
\mathcal{H}_{M\gamma} =
    \sum_{i \in \mathcal{I}_\gamma} e^{i\mathbf{k}\cdot\vec{a}_i}\, T_{MX}(\vec{a}_i); ~ \mathcal{I}_\gamma = \begin{cases} \text{odd}, & \gamma = b \\ \text{even}, & \gamma = t \end{cases},
\label{eq:H_MX}
\end{align}
where $T_{MX}(\vec{a}_i)$ is the $5\times3$ SK hopping matrix associated with the M–X bond along $\vec{a}_i$.

\textbf{Inter-sublayer X–X hopping.} The top and bottom chalcogen sublattices are coupled through direct X$_t$–X$_b$ hopping along the vectors $\vec{l}_i$ defined in Eq.~(\ref{eq:V_vecs}). The corresponding Hamiltonian block is
\begin{equation}
\mathcal{H}_{tb} =
    \sum_{i=1}^{3} e^{i\mathbf{k}\cdot\vec{l}_i}\, T_V(\vec{l}_i),
\label{eq:H_tb}
\end{equation}
where $T_V(\vec{l}_i)$ are $3\times3$ hopping matrices whose elements are determined by the SK parameters of the corresponding X–X bond (see Appendix~\ref{app:matrices}). The explicit forms of $T_M$, $T_X$, $T_{MX}$, and $T_V$ are given in Appendix~\ref{app:matrices}.

Spin--orbit coupling (SOC) is not included explicitly in the present TB framework, as first-principles calculations show that SOC does not qualitatively affect the low-energy dispersion or the CBM, inducing only a modest splitting of the VBM at $\Gamma$~\cite{PRM.3.074001}. The near-gap electronic structure therefore provides a consistent basis for the spin-independent quantum-geometric and optical properties considered here.

\section{Tight-Binding Parametrization}
\label{sec:model}

%-------------------------------------------------------------
\subsection{Slater-Koster Parametrization}
\label{subsection:SK}
%-------------------------------------------------------------
All hopping matrix elements are derived within the SK two-center approximation~\cite{Slater-Koster}. The model includes nearest- and next-nearest-neighbor M–M and X–X hoppings, together with nearest-neighbor M–X coupling. The corresponding SK parameters are $V_{pd\sigma}$ and $V_{pd\pi}$ for NN M–X hopping, $V_{dd\sigma}$, $V_{dd\pi}$, and $V_{dd\delta}$ for NN M–M hopping, $V_{pp\sigma}$ and $V_{pp\pi}$ for NN X–X hopping, $K_{dd\sigma}$, $K_{dd\pi}$, and $K_{dd\delta}$ for NNN M–M hopping, and $K_{pp\sigma}$ and $K_{pp\pi}$ for NNN intra-sublayer X–X hopping. The inter-sublayer X–X hopping associated with the vectors $\vec{l}_i$ is described by the same $V_{pp\sigma}$ and $V_{pp\pi}$ integrals evaluated along the corresponding bond direction, yielding the amplitudes $r_0,\ldots,r_7$ (see Appendix~\ref{app:sk_params}). Next-nearest-neighbor M–X hopping is neglected owing to the large M–X separation ($\sim 5.2$~\AA). Including the five on-site energies, the model contains 17 free parameters per compound. These parameters are obtained by fitting the TB dispersion to the corresponding DFT band structure.
The direction-dependent hopping amplitudes $k_i$, $t_i$, $u_i$, $p_i$, $q_i$, and $r_i$ are analytic functions of $\theta$ and the SK parameters defined above. Their explicit expressions are given in Appendix~\ref{app:sk_params}, and the corresponding hopping matrices are listed in Appendix~\ref{app:matrices}. The optimized numerical values of the SK parameters are listed in Table~\ref{tab:params}.

Since the primary objective is to accurately describe the electronic states near the band gap, the fitting procedure is restricted to the highest valence band ($V_6$) and the lowest conduction band ($C_7$). The SK parameters are optimized using a simplex algorithm~\cite{f_wmse} by minimizing the weighted mean-squared error $f_{\mathrm{wMSE}}$ between the TB and DFT band energies, given by
\begin{equation*}
f_\text{wMSE} =
\sum_{\mathbf{k},\, n \in (V_6,C_7)}
w_n(\mathbf{k})
\!\left[\varepsilon^{\,\text{TB}}_n(\mathbf{k}) -
         \varepsilon^{\,\text{DFT}}_n(\mathbf{k})\right]^{\!2},
\label{eq:cost}
\end{equation*}
where $\varepsilon^{\,\text{TB}}_n(\mathbf{k})$ is the $n$-th eigenvalue of $\mathcal{H}_{TB}$, and $\varepsilon^{\,\text{DFT}}_n(\mathbf{k})$ is the corresponding DFT eigenvalue. 
The $\mathbf{k}$-resolved weight is defined as
\begin{equation*}
w_n(\mathbf{k}) =
\begin{cases}
5.0 & \text{at } \Gamma,\, M,\, \text{or } K, \\
3.0 & \text{near the VBM or CBM} \\
1.0 & \text{otherwise,}
\end{cases}
\end{equation*}
The highest weights are assigned to the high-symmetry points to accurately capture the band extrema. Intermediate weights are used in the vicinity of the VBM at $\Gamma$ and the CBM at M, while unit weight is assigned elsewhere. The optimized SK parameters obtained from this weighted fitting procedure are listed in Table~\ref{tab:params}.
\begin{table}[h]
\caption{Optimized SK TB parameters (in eV) for monolayer $\mathrm{ZrS}_2$ and $\mathrm{HfS}_2$. The parameters were obtained by independently fitting the highest occupied ($V_6$) and lowest unoccupied ($C_7$) bands of each compound to the corresponding DFT band structure.}
\label{tab:params}
\begin{tabular}{crr}
\toprule
Parameter & $\mathrm{ZrS}_2$ &  $\mathrm{HfS}_2$ \\
\midrule
  $E_{d_0}$          & $-1.5007$     & $-1.1472$   \\
  $E_{d_1}$          & $-2.2542$     & $-2.0304$   \\
  $E_{d_2}$          & $1.6748$      & $1.6231$   \\
  $E_{p_1}$          & $-5.3555$     & $-5.6310$   \\
  $E_{p_2}$          & $-4.5714$     & $-4.7654$   \\
  $V_{pp\sigma}$     & $0.9063$      & $1.0371$   \\
  $V_{pp\pi}$        & $-0.1974$     & $-0.2387$   \\
  $V_{dd\sigma}$     & $0.0188$      & $0.1490$   \\
  $V_{dd\pi}$        & $0.1288$      & $0.1999$   \\
  $V_{dd\delta}$     & $-0.1255$     & $-0.1820$   \\
  $V_{pd\sigma}$     & $-1.4874$     & $-1.6199$   \\
  $V_{pd\pi}$        & $0.9276$      & $0.9885$   \\
  $K_{dd\sigma}$     & $-0.1167$     & $-0.2041$   \\
  $K_{dd\pi}$        & $-0.0165$     & $-0.0082$   \\
  $K_{dd\delta}$     & $0.0605$      & $0.0909$   \\
  $K_{pp\sigma}$     & $-0.0078$     & $0.0122$   \\
  $K_{pp\pi}$        & $-0.0109$     & $-0.0205$   \\
\bottomrule
\end{tabular}
\end{table}

%-------------------------------------------------------------
\subsection{Band structure and orbital character}
\label{subsec:bands}
%-------------------------------------------------------------

\begin{figure}[h]
\centering
\includegraphics[width=\columnwidth]{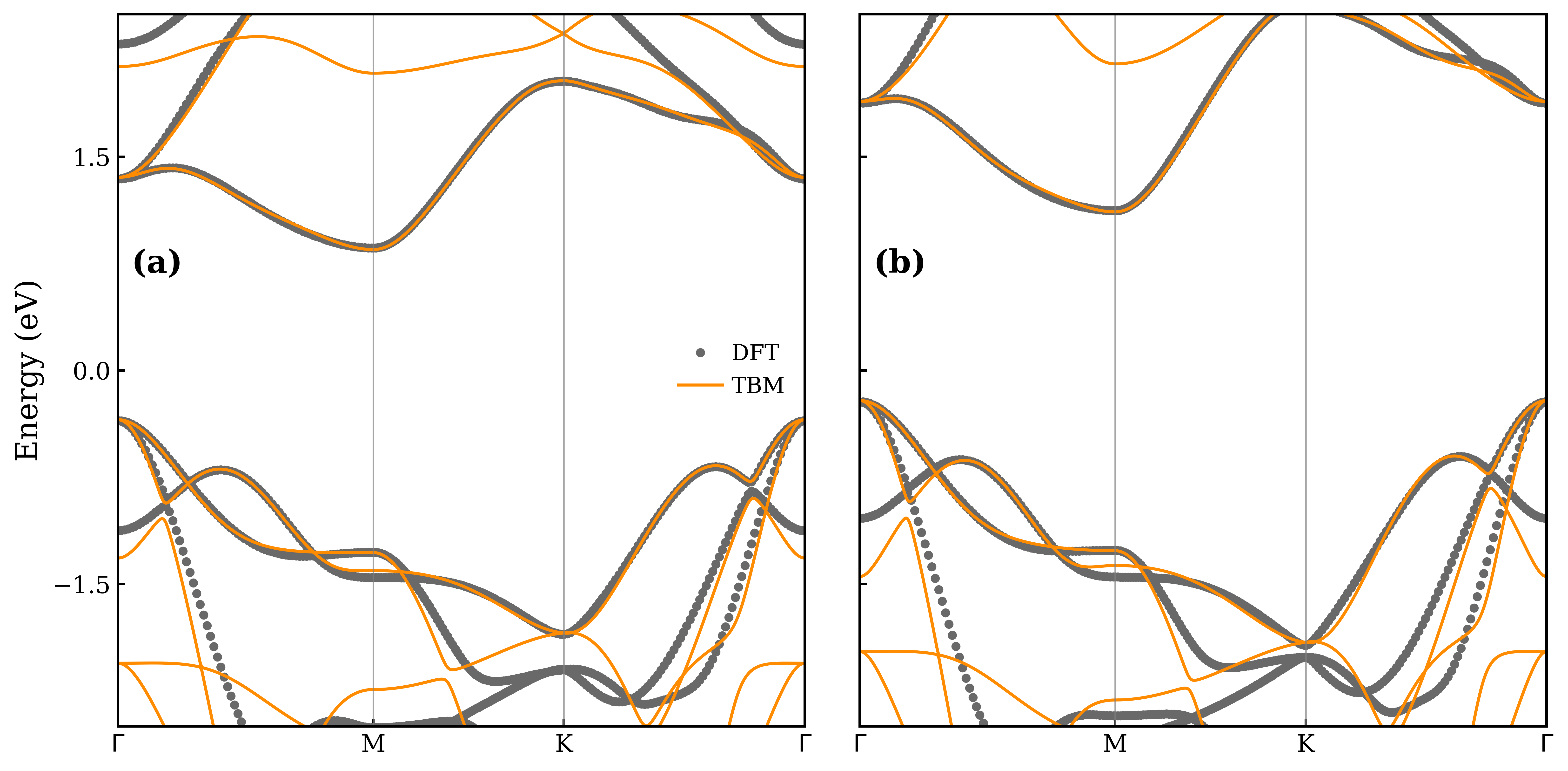}
\caption{Band structures of monolayer (a) $\mathrm{ZrS}_2$ and (b) $\mathrm{HfS}_2$: DFT (gray dots) and TB model (orange solid lines).}
\label{fig:bands}
\end{figure}

The optimized parameters exhibit similar trends for both $\mathrm{ZrS}_2$ and $\mathrm{HfS}_2$. In particular, the nearest-neighbor M–X hopping amplitudes are significantly larger than the M–M and X–X hoppings, highlighting the dominant role of metal–chalcogen hybridization. The next-nearest-neighbor corrections remain comparatively small, indicating that the electronic structure is primarily governed by short-range interactions.
\begin{table}[htbp]
\caption{Orbital contributions at the valence  and conduction band edges of monolayer $\mathrm{ZrS}_2$ and $\mathrm{HfS}_2$ obtained from DFT and the TB model. $\Gamma^{v}$ ($\Gamma^{c}$) and M$^{v}$ (M$^{c}$) denote the valence  and conduction band edges at $\Gamma$ and M, respectively. All orbital weights are normalized to unity within each column.
}
\label{tab:orbital_char}
\begin{tabular}{llcccccccc}
  \toprule
  & &
  \multicolumn{2}{c}{$\Gamma^{v}$} &
  \multicolumn{2}{c}{$\Gamma^{c}$} &
  \multicolumn{2}{c}{$\mathrm{M}^{v}$} &
  \multicolumn{2}{c}{$\mathrm{M}^{c}$} \\
  \midrule
  & & \text{DFT} & \text{TB}
    & \text{DFT} & \text{TB}
    & \text{DFT} & \text{TB}
    & \text{DFT} & \text{TB} \\
  \midrule
  \multirow{5}{*}{$\mathrm{ZrS}_2$}
    & $d_0$    & 0.00 & 0.00 & 0.00 & 0.00 & 0.00 & 0.00 & 0.34 & 0.36 \\
    & $d_1$    & 0.00 & 0.00 & 0.66 & 0.82 & 0.00 & 0.00 & 0.46 & 0.57 \\
    & $d_2$    & 0.00 & 0.00 & 0.32 & 0.08 & 0.00 & 0.00 & 0.04 & 0.07 \\
    & $p_1$    & 1.00 & 1.00 & 0.02 & 0.10 & 1.00 & 1.00 & 0.01 & 0.00 \\
    & $p_2$    & 0.00 & 0.00 & 0.00 & 0.00 & 0.00 & 0.00 & 0.09 & 0.02 \\
  \midrule
  \multirow{5}{*}{$\mathrm{HfS}_2$}
    & $d_0$    & 0.00 & 0.00 & 0.00 & 0.00 & 0.00 & 0.00 & 0.30 & 0.33 \\
    & $d_1$    & 0.00 & 0.00 & 0.66 & 0.82 & 0.00 & 0.00 & 0.36 & 0.57 \\
    & $d_2$    & 0.00 & 0.00 & 0.32 & 0.04 & 0.00 & 0.00 & 0.16 & 0.00 \\
    & $p_1$    & 1.00 & 1.00 & 0.02 & 0.14 & 1.00 & 1.00 & 0.02 & 0.00 \\
    & $p_2$    & 0.00 & 0.00 & 0.00 & 0.00 & 0.00 & 0.00 & 0.16 & 0.10 \\
  \bottomrule
\end{tabular}
\end{table}
The TB model captures the DFT band dispersion along the $\Gamma$--M--K--$\Gamma$ path and yields excellent agreement with the DFT band gaps for both compounds, which are also consistent with previously reported values in literature (Table~\ref{tab:gaps} and Fig.~\ref{fig:bands}). The inclusion of next-nearest-neighbor hopping terms significantly improves the description of the conduction band dispersion near the M point~\cite{Dias2018, Silva-Guillen2016}.

The orbital character at the band edges is summarized in Table~\ref{tab:orbital_char}. For both $\mathrm{ZrS}_2$ and $\mathrm{HfS}_2$, the TB model reproduces the dominant orbital composition of the valence  and conduction band extrema obtained from DFT. In particular, the VBM at $\Gamma$ is derived primarily from chalcogen $p_{x,y}$ states, while the CBM at M is dominated by metal $d$ orbitals in both descriptions.

%=============================================================
\section{Electronic and Optical Properties}
\label{sec:electronic_optical}

With the Slater-Koster parameters in Table~\ref{tab:params} established, we now examine the electronic and optical properties arising from the TB model.
%-------------------------------------------------------------
\subsection{Density of states}
\label{subsec:dos}
%-------------------------------------------------------------

\begin{figure}[ht!]
\centering
\includegraphics[width=\columnwidth]{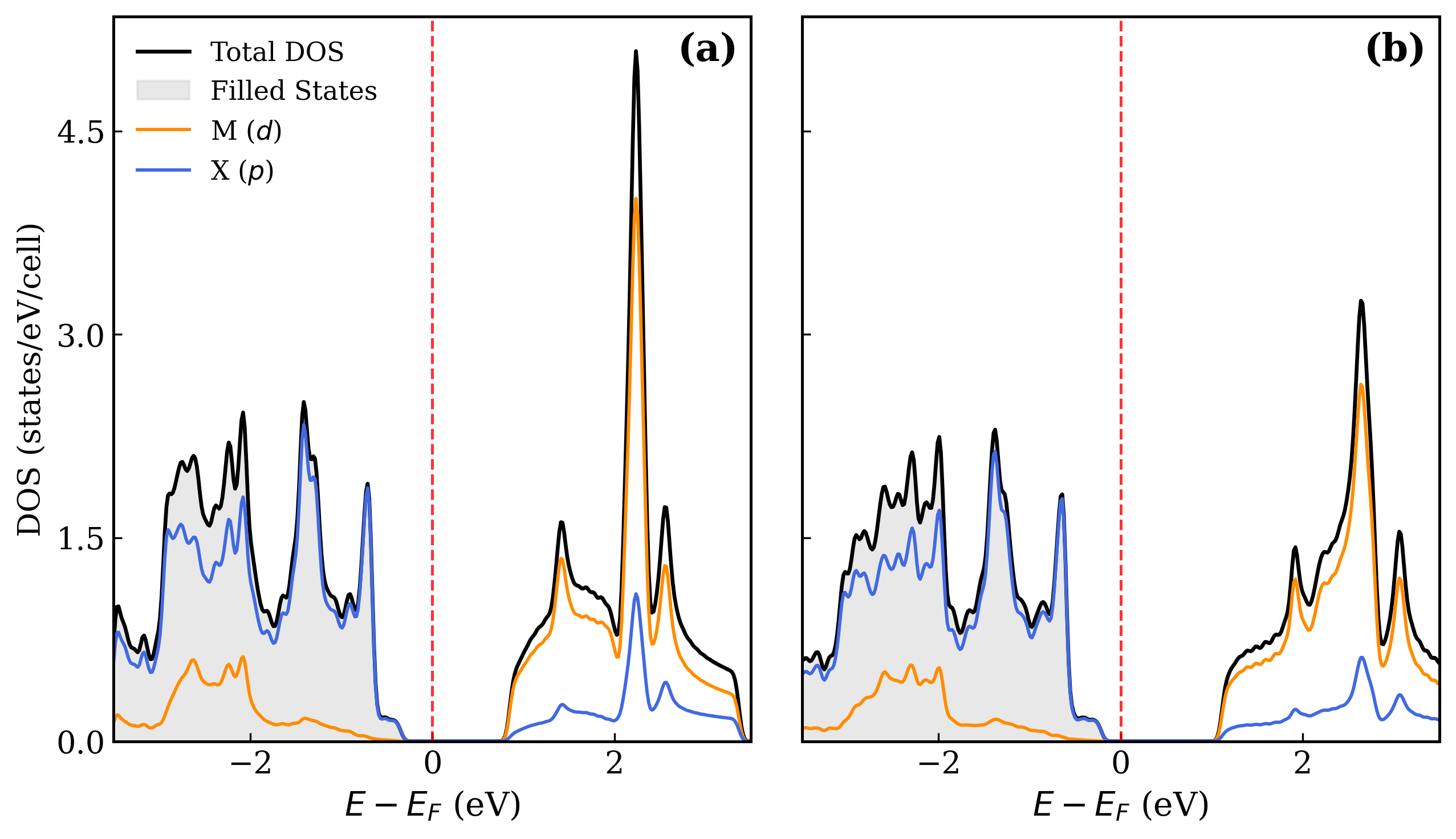}
\caption{Total and orbital-projected DOS for monolayer (a) $\mathrm{ZrS}_2$ and (b) $\mathrm{HfS}_2$ obtained from the TB model. The shaded region indicates occupied states, and the red dashed line marks the Fermi level.}
\label{fig:dos}
\end{figure}

The total and orbital-projected densities of states (DOS) for monolayer $\mathrm{ZrS}_2$ and $\mathrm{HfS}_2$ are shown in Fig.~\ref{fig:dos}. In both compounds, the valence bands are dominated by chalcogen $p$ states, whereas the conduction bands are primarily composed of transition-metal $d$ states. The clear separation between the dominant orbital characters on either side of the band gap indicates a charge transfer insulating state, in agreement with the orbital-resolved band structures shown in Fig.~\ref{fig:bands}.

The projected DOS further reveals strong metal-chalcogen hybridization throughout the valence and conduction manifolds, although the dominant orbital character remains unchanged across the gap. The conduction band region is characterized by pronounced $d$-derived peaks, reflecting the relatively weak dispersion of several conduction bands. Similar features are observed for both $\mathrm{ZrS}_2$ and $\mathrm{HfS}_2$, indicating that the low-energy electronic structure is governed by the same underlying crystal-field environment and metal-chalcogen bonding mechanism.

Overall, the DOS analysis corroborates the orbital character obtained from the band structure calculations and confirms that the essential electronic features of both compounds are accurately captured by the TB model. While the DOS establishes the orbital character of the electronic states, it does not directly quantify their dispersion near the band extrema. We therefore next examine the curvature of the band edges through the corresponding effective masses.

%-------------------------------------------------------------
\subsection{Effective masses}
\label{subsec:effmass}
%-------------------------------------------------------------
Effective masses obtained from parabolic fits to the TB bands near the valence- and conduction-band extrema are summarized in Table~\ref{tab:effmass}. The VBM at $\Gamma$ is nearly isotropic in both compounds, consistent with the threefold rotational symmetry of the lattice. In contrast, the CBM at $M$ exhibits a pronounced anisotropy, with a substantially larger effective mass along $\Gamma$--$M$ than along $M$--$K$. This anisotropy reflects the direction-dependent dispersion of the metal-$d$ states that dominate the conduction-band edge [Table~\ref{tab:orbital_char}] and is expected to give rise to anisotropic electron transport near the band minimum.

The effective masses, however, characterize only the local curvature of the bands in the vicinity of $\Gamma$ and $M$. Optical transitions, in contrast, involve electronic states throughout the Brillouin zone and therefore depend on the full momentum-space distribution of available transitions and their associated matrix elements. We first examine the joint density of states (JDOS) to identify the dominant energy ranges for interband transitions and subsequently analyze their contributions to the optical conductivity.

\begin{table}[htbp!]
\centering
\caption{Effective masses (in units of $m_e$) for $\mathrm{ZrS}_2$ and $\mathrm{HfS}_2$.}
\label{tab:effmass}
\begin{tabular}{@{}lccccccc@{}}
\toprule
\multirow{2}{*}{Material} & \multicolumn{2}{c}{VBM $m^*$} & \multicolumn{2}{c}{CBM $m^*$} & \multirow{2}{*}{$m_{\mathrm{dos}}^{\mathrm{VBM}}$} & \multirow{2}{*}{$m_{\mathrm{dos}}^{\mathrm{CBM}}$} \\
\cmidrule(lr){2-3} \cmidrule(lr){4-5}
& $\Gamma\!\to\!M$ & $\Gamma\!\to\!K$ & $M\!\to\!\Gamma$ & $M\!\to\!K$ & & & \\
\midrule
$\mathrm{ZrS}_2$  & 0.425 & 0.425 & 1.289 & 0.289 & 0.425 & 0.610 \\
$\mathrm{HfS}_2$  & 0.462 & 0.462 & 1.277 & 0.254 & 0.462 & 0.570 \\
\bottomrule
\end{tabular}
\end{table}

%-------------------------------------------------------------
\subsection{Joint Density of States}
\label{subsec:jdos}
\begin{figure}[h]
\centering
\includegraphics[width=\columnwidth]{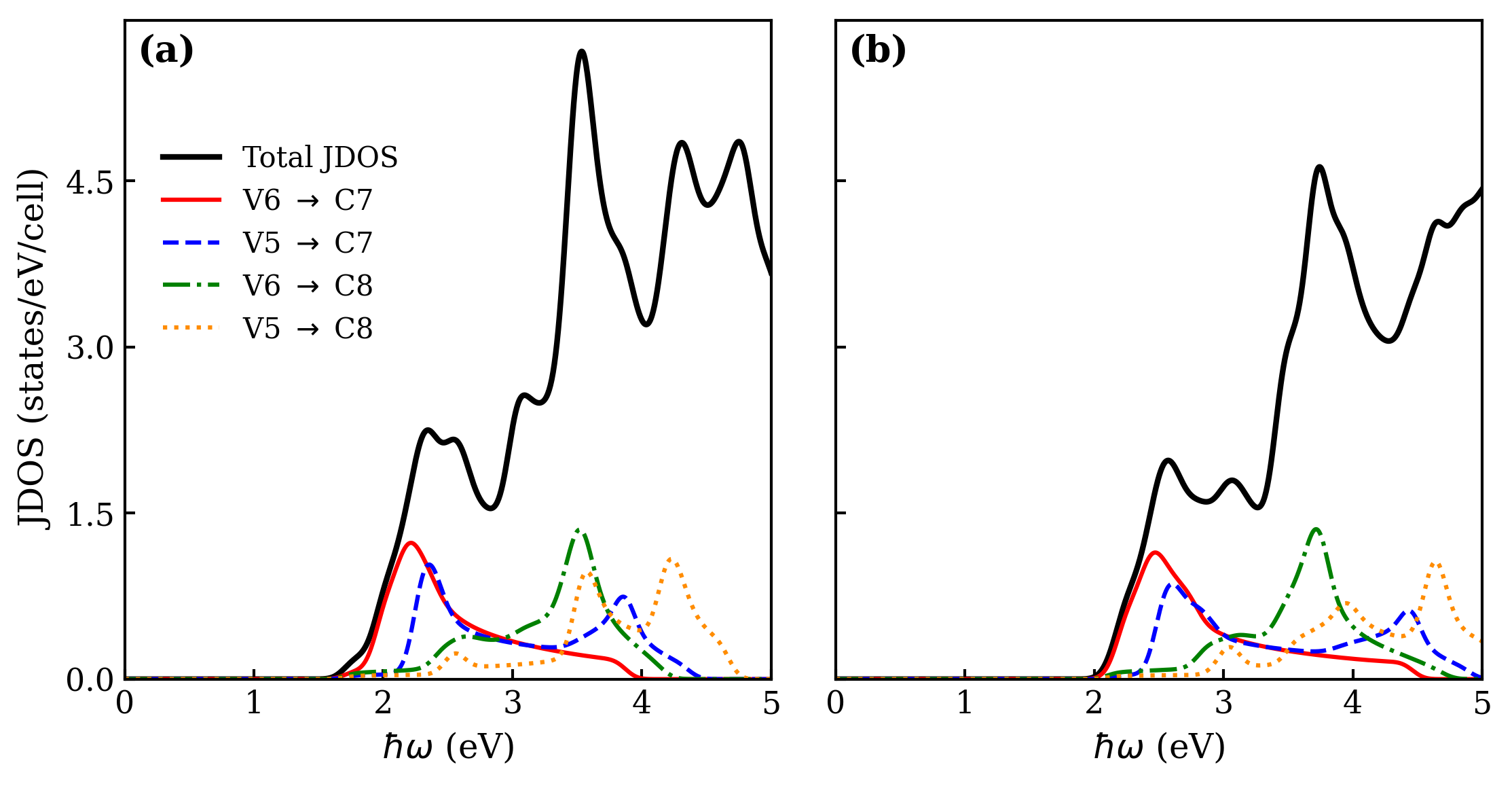}
\caption{JDOS for monolayer (a) $\mathrm{ZrS}_2$ and (b) $\mathrm{HfS}_2$. The black curves represent the total JDOS, and the colored curves show the leading band-pair contributions arising from transitions between the upper valence and lower conduction bands.}
\label{fig:jdos}
\end{figure}

%-------------------------------------------------------------
The joint density of states (JDOS), shown in Fig.~\ref{fig:jdos}, quantifies the density of available interband transitions at a given photon energy without weighting them by their optical matrix elements. We label the occupied and unoccupied states as $V_i$ ($i=1,\ldots,6$) and $C_i$ ($i=7,\ldots,11$), respectively, according to their energy ordering in the eleven-band Hamiltonian, such that $V_6$ and $C_7$ denote the VBM and CBM. For both compounds, the JDOS increases with photon energy as additional valence-to-conduction channels become accessible, with the overall similarity between $\mathrm{ZrS}_2$ and $\mathrm{HfS}_2$ reflecting their closely related electronic structures. The prominent low energy peak is primarily associated with the near-degenerate $V_6\rightarrow C_7$ and $V_5\rightarrow C_7$ transitions, while the higher energy shoulder is dominated by the $V_6\rightarrow C_8$ and $V_5\rightarrow C_8$ channels.

Having identified the available interband channels, we next examine the optical conductivity, where each transition is weighted by its corresponding interband velocity matrix element.

%-------------------------------------------------------------
\subsection{Optical conductivity}
\label{subsec:opt_results}
%-------------------------------------------------------------
The frequency-dependent optical conductivity tensor for a non-interacting system computed within the Kubo-Greenwood linear-response formalism~\cite{kubo1957,Greenwood_1958} is given by
\begin{equation}
\sigma_{\alpha\beta}(\omega)
  \!=\! \frac{e^2}{i\hbar A}
    \!\sum_{\mathbf{k},m,n} \!
    \frac{f_{mn}}
         {\Delta\epsilon_{mn}}
    \frac{v_{\mathbf{k},\alpha}^{mn}\,v_{\mathbf{k},\beta}^{nm}}
         {\hbar\omega+\Delta\epsilon_{mn}+i\eta},
\label{eq:kubo_full}
\end{equation}
where $\alpha,\beta\in{x,y,z}$, $A$ is the unit-cell area, and $\eta$ is the broadening parameter. Now onward, explicit dependence on $\omega$ is omitted for brevity. Here, $f_{mn}=f(\epsilon_{\mathbf{k},m})-f(\epsilon_{\mathbf{k},n})$ with $f(\epsilon_k)$ denoting the Fermi-Dirac distribution, $\Delta\epsilon_{mn}=\epsilon_{\mathbf{k},m}-\epsilon_{\mathbf{k},n}$ and $v_{\mathbf{k},\alpha}^{mn}=\langle m_{\mathbf{k}}|\hat{v}_{\alpha}|n_{\mathbf{k}}\rangle$ are the velocity matrix elements with $\hat{v}_{\alpha}=\frac{1}{\hbar}\partial_{\alpha}\mathcal{H}$. For the undoped 1T-$\mathrm{MX}_2$ monolayers considered here, the Fermi level lies within the band gap, with fully occupied valence and empty conduction bands, so the intraband (Drude) contribution vanishes, and the optical response is entirely interband.

Following the Ref.~\cite{Torma2023}, the interband conductivity tensor is decomposed into symmetric ($s$) and antisymmetric ($a$) parts with respect to band-index exchange:
\begin{equation}
\sigma^{\mathrm{inter}}_{\alpha\beta}
  = \sigma^{s}_{\alpha\beta}
  + \sigma^{a}_{\alpha\beta}.
\label{eq:s_plus_a}
\end{equation}
For the diagonal components ($\alpha=\beta$), only the symmetric contribution survives, whose real part in the limit $\eta\rightarrow0^+$ yields
\begin{align}
\operatorname{Re}\sigma_{\alpha\alpha} \!&=\! \frac{\pi e^2}{A} \!\sum_{\mathbf{k},n\neq m}\!
    \frac{f_{mn}} {\Delta\epsilon_{mn}} \bigl|v^{mn}_{\mathbf{k},\alpha}\bigr|^2
    \delta \bigl(\hbar\omega-\Delta\epsilon_{mn}\bigr).
\label{eq:re_sigma_xx}
\end{align}
For $m\neq n$, the interband velocity matrix element is related to the derivatives of the Bloch states by
\begin{equation}
v_{\mathbf{k},\alpha}^{mn} = \Delta\epsilon_{mn} \braket{\partial_{k_{\alpha}}m_\mathbf{k} | n_\mathbf{k}}
\label{eq:velocity_berry}
\end{equation}
Thus, the interband optical response is directly expressed in terms of interband Bloch-state derivative matrix elements. The imaginary part can be obtained directly from $\mathrm{Re}\,\sigma_{\alpha\alpha}$ through the Kramers--Kr\"onig relation. For the off-diagonal components ($\alpha \neq \beta$), the product $v^{mn}_{\alpha}v^{nm}_{\beta}$ is generally complex, giving rise to both symmetric and antisymmetric contributions to the optical conductivity. Within the quantum geometric decomposition, these contributions are associated with the quantum metric and Berry curvature, respectively~\cite{Ahn2022,Torma2023,Ghosh2024}. Since the presence of combined time-reversal and inversion symmetries in pristine 1T-$\mathrm{MX}_2$ monolayers forces the Berry curvature to be zero throughout the Brillouin zone~\cite{Xiao2010,Nagaosa2010}, the antisymmetric conductivity vanishes at all frequencies, leaving the quantum metric as the sole geometric contribution to the linear optical response. Furthermore, the threefold rotational symmetry of the $D_{3d}$ point group imposes stringent constraints on the optical conductivity tensor, forcing the symmetric off-diagonal component $\sigma_{xy}$ to vanish. Consequently, only the diagonal components remain finite and are related by  $\sigma_{xx}=\sigma_{yy}$.

\begin{figure}[h]
\centering
\includegraphics[width=\columnwidth]{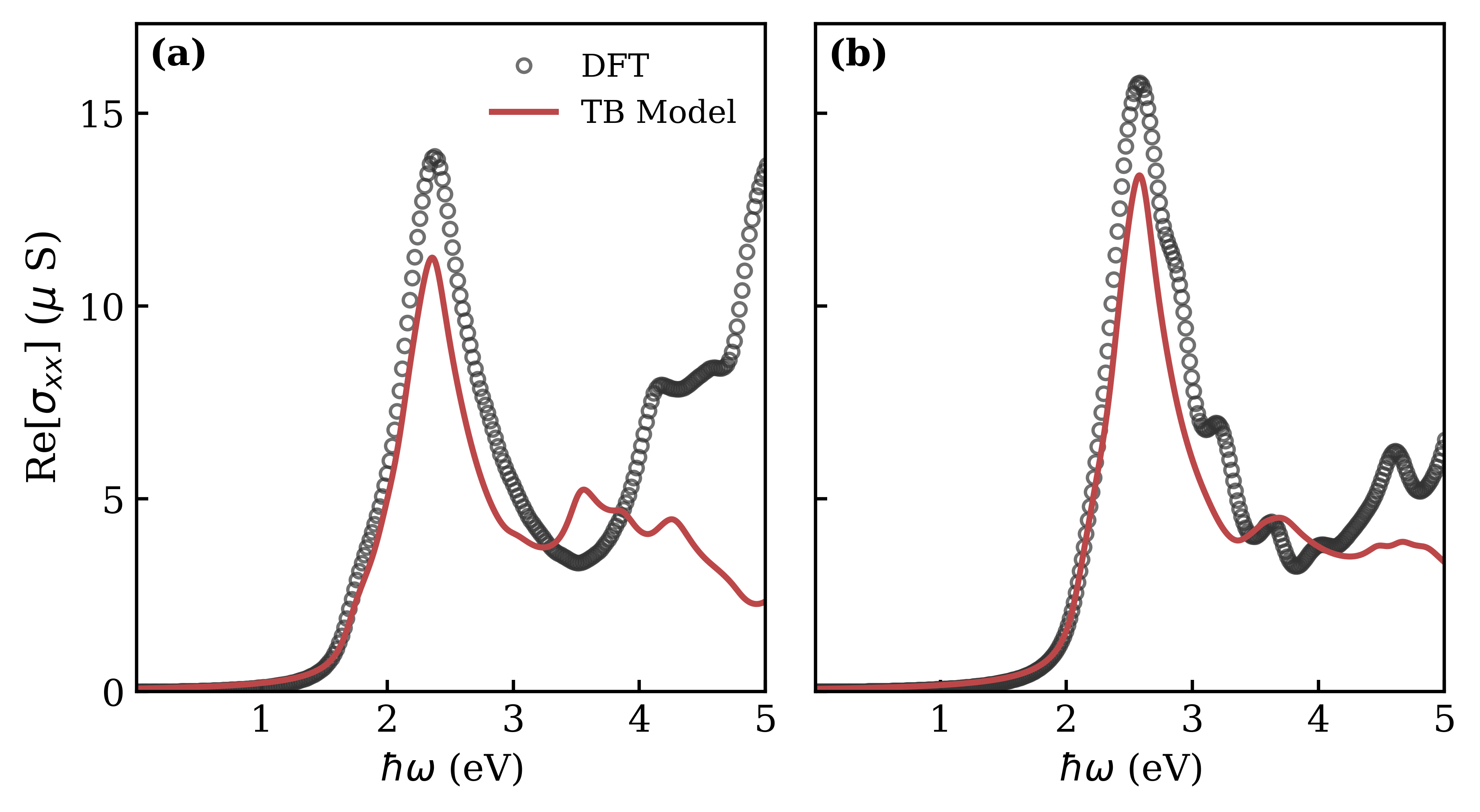}
\caption{Comparison of the real part of the optical conductivity $\mathrm{Re}\,\sigma_{xx}$, obtained from DFT (gray dots) and the TB model (red solid lines) for monolayer (a) $\mathrm{ZrS}_2$ and (b) $\mathrm{HfS}_2$.
}
\label{fig:opt_dft_tb}
\end{figure}

Figure~\ref{fig:opt_dft_tb} shows the real part of the optical conductivity, $\mathrm{Re}\,\sigma_{xx}$, calculated using the TB model with a Lorentzian broadening of $\eta=0.1$ eV, alongside the corresponding DFT results for comparison. For both the compounds, the TB and DFT calculations exhibit a consistent low-energy optical response, with similar absorption onsets and dominant spectral features occurring at closely related energies. The JDOS analysis in Fig.~\ref{fig:jdos} identifies this low-energy feature primarily with the near-degenerate $V_6\rightarrow C_7$ and $V_5\rightarrow C_7$ transitions. These transitions involve the near-gap bands explicitly prioritized in the present parametrization, providing the relevant low-energy states for the optical response. Differences from DFT become more pronounced at higher photon energies, where bands farther from the gap were not explicitly prioritized in the fit. The comparison establishes that the present parametrization accurately describes the low-energy optical response targeted by the model.

For the in-plane response, we write $\varepsilon_{xx} = \varepsilon_{xx}^{(1)} +i\varepsilon_{xx}^{(2)}$ with $\varepsilon_{xx}=\varepsilon_{yy}$ by symmetry. The dielectric function is related to the conductivity by
\begin{equation*}
\varepsilon^{(1)}_{xx} = 1-\frac{\operatorname{Im}\sigma_{xx}}{\varepsilon_0\omega},
\qquad
\varepsilon^{(2)}_{xx} = \frac{\operatorname{Re}\sigma_{xx}}{\varepsilon_0\omega}.
\label{eq:dielectric_conductivity}
\end{equation*}

\begin{figure}[h]
\centering
\includegraphics[width=\columnwidth]{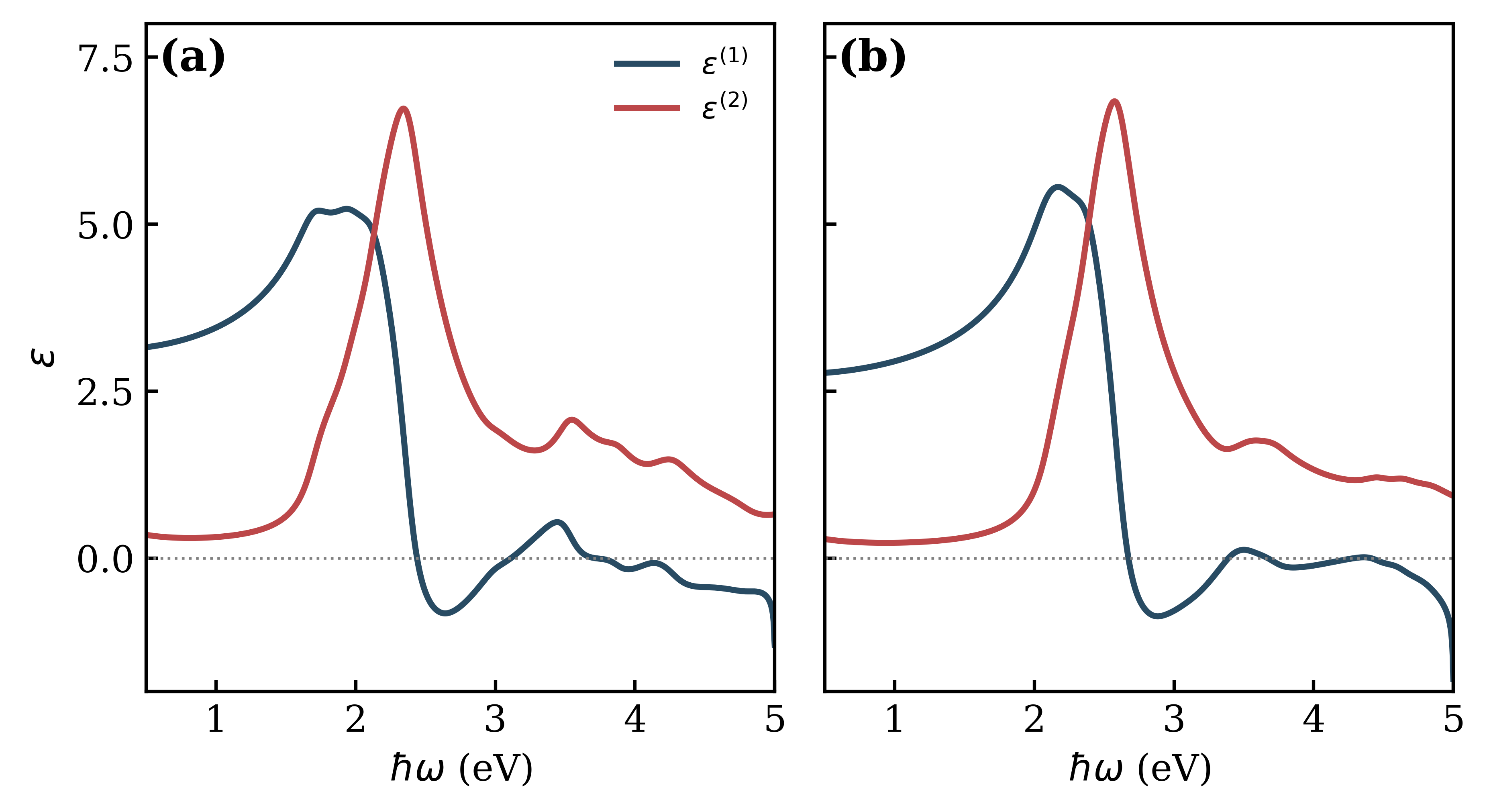}
\caption{Real ($\varepsilon^{(1)}$) and imaginary ($\varepsilon^{(2)}$) parts of the frequency-dependent dielectric function for monolayer (a) $\mathrm{ZrS}_2$ and (b) $\mathrm{HfS}_2$ obtained from the optical conductivity.
}
\label{fig:dielectric}
\end{figure}

Fig.~\ref{fig:dielectric} shows the resulting dielectric functions. For simplicity, we suppress the $xx$ indices on $\varepsilon$. Both compounds exhibit a pronounced $\varepsilon^{(2)}$ peak in the visible--UV range, accompanied by a rise of $\varepsilon^{(1)}$ near the absorption onset followed by a decrease at higher energies. $\mathrm{HfS}_2$ exhibits a modestly taller $\varepsilon^{(2)}$ peak shifted to higher energy relative to $\mathrm{ZrS}_2$, consistent with its larger $V_{pd\sigma}$ and $V_{pd\pi}$ hopping amplitudes (Table~\ref{tab:params}), which enhance metal--chalcogen hybridization and can redistribute the optical weight toward a narrower energy range.

\begin{figure*}[htbp]
\centering
\includegraphics[width=0.8\textwidth]{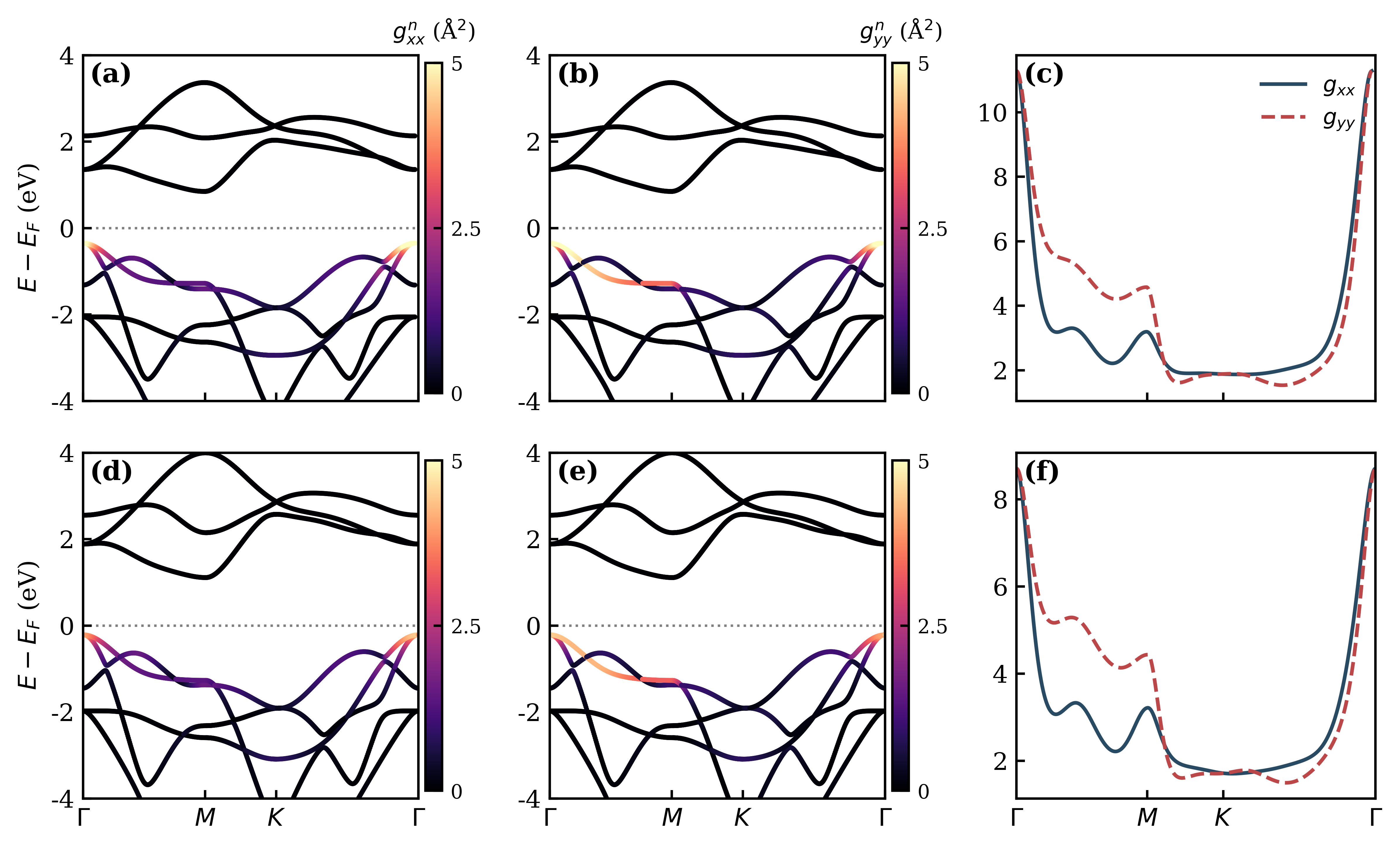}
\caption{Band-resolved quantum metric for $\mathrm{ZrS}_2$ (top row) and $\mathrm{HfS}_2$ (bottom row) along $\Gamma$--$M$--$K$--$\Gamma$. The left and middle columns show the bands colored by $g^n_{xx}(\mathbf{k})$ and $g^n_{yy}(\mathbf{k})$, respectively, while the right column shows the corresponding occupied-state metrics $g_{xx}(\mathbf{k})=\sum_n f(\epsilon_{\mathbf{k},n})g^n_{xx}(\mathbf{k})$ and $g_{yy}(\mathbf{k})$.
}
\label{fig:qm_bands}
\end{figure*}

The interband $f$-sum rule establishes a direct connection between the optical conductivity and the BZ-averaged quantum metric~\cite{Souza2000,Torma2023}. For the diagonal conductivity, it takes the form
\begin{equation}
\int_0^{\infty}
\frac{\operatorname{Re} \sigma_{xx}}{\omega} d\omega = \frac{\pi e^2}{\hbar} G_{xx}, 
\label{eq:fsum}
\end{equation}
where $G_{xx}$ is $xx$-component of the BZ-averaged occupation-weighted quantum metric $G_{\alpha\beta}$ and is defined as
\begin{equation}
G_{\alpha\beta}
= \frac{1}{A_{\mathrm{BZ}}}
\int_{\mathrm{BZ}} \sum_n f(\epsilon_{\mathbf{k},n})
g^n_{\alpha\beta}d^2k.
\label{eq:G_integrated}
\end{equation}
Here, the local quantum metric $g^n_{\alpha\beta}$ is evaluated using the gauge-invariant expression~\cite{Provost1980,Resta2011} in which the summation is restricted to occupied-to-unoccupied transitions to ensure a gauge-invariant representation of the occupied state geometry:
\begin{equation}
g^n_{\alpha\beta}
  = \operatorname{Re}\!\sum_{\substack{m \\ \epsilon_m>\epsilon_F}}
    \frac{\langle n_\mathbf{k}|\partial_{k_\alpha}\mathcal{H}|m_\mathbf{k}\rangle
          \langle m_\mathbf{k}|\partial_{k_\beta}\mathcal{H}|n_\mathbf{k}\rangle}
         {(\Delta\epsilon_{nm})^2}.
\label{eq:gn_kubo}
\end{equation}
The band-resolved decomposition in Fig.~\ref{fig:qm_bands} shows that the quantum metric is dominated by the two highest occupied bands, $V_5$ and $V_6$, which together account for $\sim78\%$ of the integrated $G_{xx}$ in both compounds. The remaining contribution is distributed between $V_4$ ($\sim10\%$) and the three lower occupied bands ($\sim12\%$). This dominance reflects the small energy separation of $V_5$ and $V_6$ from the lowest conduction band, $C_7$, together with the large interband velocity matrix elements associated with the $p$--$d$ hybridization at the band edges (Table~\ref{tab:orbital_char}). The corresponding momentum-resolved distribution of the occupied-state metrics $g_{xx}$ and $g_{yy}$ (shown in the right-hand panels of Fig.~\ref{fig:qm_bands}), exhibit pronounced momentum dependence with enhanced weight near $\Gamma$. Their near identical distributions in both compounds confirm that the quantum metric is governed primarily by the shared near-gap electronic structure rather than by the specific transition-metal species.

Substituting this local quantum metric into Eq.~\eqref{eq:G_integrated} and performing the Brillouin-zone integration, we obtain $G_{xx}=2.559$~\AA$^2$ for $\mathrm{ZrS}_2$ and $G_{xx}=2.470$~\AA$^2$ for $\mathrm{HfS}_2$. The resulting quantum metric tensor satisfies $G_{xx}=G_{yy}$ and $G_{xy}=0$, as required by the $D_{3d}$ symmetry of the monolayer 1T TMDs~\cite{Ozawa2018}.
\begin{table}[h]
\caption{Verification of the interband $f$-sum rule. Comparison between the quantum metric prediction $\pi G_{xx}$ and the optical spectral weight $W_{\mathrm{opt}}=\int_0^{\infty}
\operatorname{Re} \sigma_{xx}/\omega ~d\omega$ obtained from the interband optical conductivity.}
\label{tab:fsum}
\begin{tabular}{lccc}
\toprule
Compound & $\pi G_{xx}$(\AA$^2$) & $W_{\rm opt}$ (\AA$^2$) & Deviation (\%) \\
\midrule
$\mathrm{ZrS}_2$ & 8.041 & 8.055 & 0.2 \\
$\mathrm{HfS}_2$ & 7.760 & 7.770 & 0.1 \\
\bottomrule
\end{tabular}
\end{table}
To verify Eq.~\eqref{eq:fsum}, both sides were evaluated independently using $\eta=0.001$~eV, which closely approaches the sharp-transition limit while avoiding numerical singularities. The results, summarized in Table~\ref{tab:fsum}, agree to within $0.2\%$ for $\mathrm{ZrS}_2$ and $0.1\%$ for $\mathrm{HfS}_2$. For the larger $\eta=0.1$~eV used for the optical spectra in Sec.~\ref{subsec:opt_results}, the deviation increases to $\sim2\%$ but decreases systematically as $\eta$ is reduced toward zero. This behavior demonstrates that the residual deviation at finite broadening arises from the Lorentzian broadening, while the interband $f$-sum rule is recovered in the sharp-transition limit. This fulfillment of the interband $f$-sum rule establishes a direct quantitative connection between the optical spectral weight and the quantum metric of the occupied electronic states. Consequently, the integrated optical spectral weight provides an experimentally accessible probe of the underlying quantum geometry, linking measurable optical observables to the geometric properties of Bloch states.

%=============================================================
\section{Conclusion}
\label{sec:conclusion}
%=============================================================
In this work, we have developed a generalized eleven-band tight-binding Hamiltonian within the Slater--Koster framework for monolayer 1T-$\mathrm{MX}_2$ compounds. The model incorporates the five transition-metal $d$ orbitals and six chalcogen $p$ orbitals in an orthogonal basis, with hopping parameters constrained by the crystal geometry and lattice symmetries. As representative examples, we have constructed parametrizations for $\mathrm{ZrS}_2$ and $\mathrm{HfS}_2$, which reproduce the key features of their low-energy electronic structures obtained from DFT, including the indirect band gaps, orbital character, and crystal-field splittings. The good agreement with the DFT band structures establishes the eleven-band model as a reliable and physically transparent platform for investigating the microscopic electronic and geometric properties of this class of materials.

Using the resulting Hamiltonian, we have investigated the quantum geometry of the occupied Bloch states. Owing to the simultaneous presence of time-reversal and inversion symmetries, the Berry curvature, corresponding to the imaginary part of the quantum geometric tensor, vanishes identically, whereas the quantum metric, given by its real part, remains finite. We find that the quantum metric originates from interband mixing between occupied and unoccupied states and becomes particularly pronounced near the band edges, where the relatively small energy separation is accompanied by substantial $p$--$d$ hybridization. Remarkably, the quantum-metric contribution is strongly concentrated in the two highest occupied bands, V$_5$ and V$_6$, which together account for approximately 78$\%$ of the integrated G$_{xx}$ in both $\mathrm{ZrS}_2$ and $\mathrm{HfS}_2$. This band-resolved analysis provides a direct microscopic picture of the origin of the quantum geometry in these materials. We have further established a direct connection between the quantum metric and the optical response by calculating the optical conductivity within the Kubo formalism. In both compounds, we verify the interband $f$-sum rule, demonstrating that the integrated optical spectral weight is directly related to the Brillouin-zone-averaged quantum metric. This establishes optical conductivity as an experimentally accessible probe of the quantum geometry of occupied Bloch states. The connection is particularly significant in the present centrosymmetric and time-reversal-invariant systems, where symmetry enforces a vanishing Berry curvature, thereby isolating the quantum metric as the relevant geometric contribution to the optical response.

Overall, our results establish an eleven-band Slater--Koster framework that simultaneously provides an accurate description of the electronic structure and a microscopic understanding of quantum-geometric effects in monolayer 1T-$\mathrm{MX}_2$ compounds. As a natural extension of the present work, it would be interesting to investigate how the quantum geometry evolves under controlled perturbations. Strain, external electric fields, and other symmetry-breaking perturbations can modify the underlying electronic structure and crystal symmetries, potentially inducing a finite Berry curvature and redistributing the quantum-metric weight among different bands and regions of the Brillouin zone. A systematic investigation of these effects, and their corresponding signatures in optical and other response functions, is beyond the scope of the present work and is left for future studies.

\section*{Acknowledgments}
We acknowledge the computing resources of `PARAM SHAVAK' at Computational Condensed Matter Physics Lab, Department of Physics, NIT Silchar. S.~N. acknowledges financial support from Anusandhan National Research Foundation (ANRF), Government of India via the Prime Minister's Early Career Research Grant: ANRF/ECRG/2024/005947/PMS. We thank Emmanuele Cappelluti for insightful discussions and Alexandre Cavalheiro Dias and Zhizi Guan for helpful suggestions.
%=============================================================
\appendix
%=============================================================

%-------------------------------------------------------------
\section{Hopping Matrices}
\label{app:matrices}
%-------------------------------------------------------------
This appendix lists the complete set of direction-dependent hopping matrices entering the TB Hamiltonian of Eq.~(\ref{eq:H_block}). All matrices are expressed in the orbital basis defined in Eq.~(\ref{eq:raw_basis}). The matrix structure, including the pattern of allowed hopping elements and their relative signs, is fixed by the crystal symmetry of the 1T phase and is therefore common to all 1T-$\mathrm{MX}_2$ compounds. Material specific differences enter only through the numerical values of the SK parameters summarized in Appendix~\ref{app:sk_params}.

For convenience, the hopping amplitudes are grouped according to their physical origin. The parameters $t_i$ denote nearest neighbor M--M hoppings, $u_i$ denote next nearest neighbor M--M hoppings, $p_i$ and $q_i$ correspond to nearest and next nearest neighbor intra-sublayer X--X hoppings, respectively, while $r_i$ describes inter-sublayer X--X coupling.

The NN intra-sublayer X--X hopping matrices $T_X(\boldsymbol{\delta}_i)$ in the $(p_x,p_y,p_z)$ basis are
\begin{align}
\
T_X(\vec{\delta}_1) &=
\begin{pmatrix} p_0 & 0 & 0 \\ 0 & p_3 & 0 \\ 0 & 0 & p_3 \end{pmatrix}
= T_X(\vec{\delta}_4), \notag
\\[2pt]
T_X(\vec{\delta}_2) &=
\begin{pmatrix} p_1 & p_2 & 0 \\ p_2 & p_4 & 0 \\ 0 & 0 & p_3 \end{pmatrix}
= T_X(\vec{\delta}_5), \notag
\\[2pt]
T_X(\vec{\delta}_3) &=
\begin{pmatrix} p_1 & -p_2 & 0 \\ -p_2 & p_4 & 0 \\ 0 & 0 & p_3 \end{pmatrix}
= T_X(\vec{\delta}_6).
\label{eq:tXS_d}
\end{align}
The nearest-neighbor M--M hopping matrices $T_M(\boldsymbol{\delta}_i)$ along the six bond directions are
\begin{align}
T_M(\vec{\delta}_1) &=
\begin{pmatrix}
t_0  & -2t_2 & 0   & 0    & 0    \\
-2t_2 & t_6  & 0   & 0    & 0    \\
0    & 0     & t_3 & 0    & 0    \\
0    & 0     & 0   & t_8  & 0    \\
0    & 0     & 0   & 0    & t_{11}
\end{pmatrix}
= T_M(\vec{\delta}_4), \notag
\\[2pt]
T_M(\vec{\delta}_2) &=
\begin{pmatrix}
t_0  & t_2  & t_1  & 0    & 0    \\
t_2  & t_7  & t_5  & 0    & 0    \\
t_1  & t_5  & t_4  & 0    & 0    \\
0    & 0    & 0    & t_9  & t_{10} \\
0    & 0    & 0    & t_{10} & t_{12}
\end{pmatrix}
= T_M(\vec{\delta}_5), \notag
\\[2pt]
T_M(\vec{\delta}_3) &=
\begin{pmatrix}
t_0  & t_2   & -t_1  & 0    & 0    \\
t_2  & t_7   & -t_5  & 0    & 0    \\
-t_1 & -t_5  & t_4   & 0    & 0    \\
0    & 0     & 0     & t_9  & -t_{10} \\
0    & 0     & 0     & -t_{10} & t_{12}
\end{pmatrix}
= T_M(\vec{\delta}_6).
\label{eq:tM_d}
\end{align}

The six NN M--X bond vectors $\mathbf{a}_i$ and the corresponding $5\times3$ hopping matrices $T_{MX}(\vec{a}_i)$ are
\begin{align}
 &T_{MX}(\vec{a}_1) &=
\begin{pmatrix}
 0 & 2k_1 & -k_2 \\
 0 & k_8 & -2k_{10} \\
 k_3 & 0 & 0 \\
 0 & -k_{11} & -2k_{13} \\
 -k_{14} & 0 & 0
\end{pmatrix}, \notag \\
& T_{MX}(\vec{a}_2) &=
\begin{pmatrix}
 k_0 & k_1 & k_2 \\
 -k_7 & -k_9 & -k_{10} \\
 -k_4 & k_5 & -k_6 \\
 -k_6 & k_{12} & -k_{13} \\
 k_{15} & -k_6 & k_{16}
\end{pmatrix}, \notag\\
&
 T_{MX}(\vec{a}_3) &=
\begin{pmatrix}
 k_0 & -k_1 & -k_2 \\
 -k_7 & k_9 & k_{10} \\
 k_4 & k_5 & -k_6 \\
 -k_6 & -k_{12} & k_{13} \\
 -k_{15} & -k_6 & k_{16}
\end{pmatrix}, \notag\\
&
 T_{MX}(\vec{a}_4) &=
\begin{pmatrix}
 0 & -2k_1 & k_2 \\
 0 & -k_8 & 2k_{10} \\
 -k_3 & 0 & 0 \\
 0 & k_{11} & 2k_{13} \\
 k_{14} & 0 & 0
\end{pmatrix}, \notag\\
&
 T_{MX}(\vec{a}_5) &=
\begin{pmatrix}
 -k_0 & -k_1 & -k_2 \\
 k_7 & k_9 & k_{10} \\
 k_4 & -k_5 & k_6 \\
 k_6 & -k_{12} & k_{13} \\
 -k_{15} & k_6 & -k_{16}
\end{pmatrix}, \notag\\
&
 T_{MX}(\vec{a}_6) &=
\begin{pmatrix}
 -k_0 & k_1 & k_2 \\
 k_7 & -k_9 & -k_{10} \\
 -k_4 & -k_5 & k_6 \\
 k_6 & k_{12} & -k_{13} \\
 k_{15} & k_6 & -k_{16}
\end{pmatrix}.
\label{eq:T_MX}
\end{align}
The inter-sublayer block is $\mathcal{H}_{tb}(\mathbf{k})
=\sum_{i=1}^{3}
e^{i\mathbf{k}\cdot\mathbf{V}_i}
T_V({l}_i)$,
with $\mathcal{H}_{bt}=\mathcal{H}_{tb}^{\dagger}$. The corresponding hopping matrices are
\begin{align}
T_V(\vec{l}_1) &=
\begin{pmatrix}
r_0 & r_2 & r_3 \\
r_2 & r_4 & r_6 \\
r_3 & r_6 & r_7
\end{pmatrix}, \notag
\\[4pt]
T_V(\vec{l}_2) &=
\begin{pmatrix}
r_1 & 0    & 0     \\
0   & r_5  & -2r_6 \\
0   & -2r_6 & r_7
\end{pmatrix}, \notag
\\[4pt]
T_V(\vec{l}_3) &=
\begin{pmatrix}
r_0  & -r_2 & -r_3 \\
-r_2 &  r_4 &  r_6 \\
-r_3 &  r_6 &  r_7
\end{pmatrix}.
\label{eq:tV}
\end{align}

The NNN intra-sublayer X--X hopping matrices along the six $\mathbf{c}_i$ directions are
\begin{align}
T_X(\vec{c}_1) &=
\begin{pmatrix} q_0 & q_2 & 0 \\ q_2 & q_3 & 0 \\ 0 & 0 & q_1 \end{pmatrix}
= T_X(\vec{c}_4), \notag
\\[2pt]
T_X(\vec{c}_2) &=
\begin{pmatrix} q_1 & 0 & 0 \\ 0 & q_4 & 0 \\ 0 & 0 & q_1 \end{pmatrix}
= T_X(\vec{c}_5), \notag
\\[2pt]
T_X(\vec{c}_3) &=
\begin{pmatrix} q_0 & -q_2 & 0 \\ -q_2 & q_3 & 0 \\ 0 & 0 & q_1 \end{pmatrix}
= T_X(\vec{c}_6).
\label{eq:tXS_c}
\end{align}
The NNN M--M hopping matrices along the six $\mathbf{c}_i$ directions are
\begin{align}
T_M(\vec{c}_1) &=
\begin{pmatrix}
u_0  & u_2  & u_1  & 0    & 0    \\
u_2  & u_6  & u_5  & 0    & 0    \\
u_1  & u_5  & u_3  & 0    & 0    \\
0    & 0    & 0    & u_8  & u_9 \\
0    & 0    & 0    & u_9 & u_{10}
\end{pmatrix}
= T_M(\vec{c}_4), \notag
\\[2pt]
T_M(\vec{c}_2) &=
\begin{pmatrix}
u_0  & -2u_2  & 0    & 0    & 0    \\
-2u_2 & u_7  & 0    & 0    & 0    \\
0    & 0    & u_4  & 0    & 0    \\
0    & 0    & 0    & u_4 & 0  \\
0    & 0    & 0    & 0    & u_{11}
\end{pmatrix}
= T_M(\vec{c}_5), \notag
\\[2pt]
T_M(\vec{c}_3) &=
\begin{pmatrix}
u_0  & u_2   & -u_1  & 0    & 0    \\
u_2  & u_6   & -u_5  & 0    & 0    \\
-u_1 & -u_5  & u_3   & 0    & 0    \\
0    & 0     & 0     & u_8  & -u_9 \\
0    & 0     & 0     & -u_9 & u_{10}
\end{pmatrix}
= T_M(\vec{c}_6).
\label{eq:tM_c}
\end{align}

%-------------------------------------------------------------
\section{Slater-Koster Expressions for the Hopping Amplitudes}
\label{app:sk_params}
%-------------------------------------------------------------
The hopping amplitudes appearing in the matrices of Appendix~\ref{app:matrices} are obtained from the standard Slater-Koster two-center rotation formalism~\cite{Slater-Koster}. Throughout this appendix, we define $c\equiv\cos\theta$ and $s\equiv\sin\theta$, where $\theta$ is the M--X bond angle listed in Table~\ref{tab:structure}. The parameters $t_i$, $u_i$, $p_i$, $q_i$, and $r_i$ denote the NN M--M, NNN M--M, NN and NNN intra-sublayer X--X, and inter-sublayer X--X hopping amplitudes, respectively. Their explicit expressions in terms of the SK integrals are listed below.
\begin{align}
p_0 &= V_{pp\sigma},\notag\\
p_1 &= \tfrac{1}{4}V_{pp\sigma} + \tfrac{3}{4}V_{pp\pi},\notag\\
p_2 &= \tfrac{\sqrt{3}}{4}(V_{pp\sigma} - V_{pp\pi}),\notag\\
p_3 &= V_{pp\pi},\notag\\
p_4 &= \tfrac{3}{4}V_{pp\sigma} + \tfrac{1}{4}V_{pp\pi}.
\label{eq:p_params}
\end{align}
\begin{align}
q_0 &= \tfrac{3}{4}K_{pp\sigma} + \tfrac{1}{4}K_{pp\pi}, \notag\\
q_1 &= K_{pp\pi}, \notag\\
q_2 &= \tfrac{\sqrt{3}}{4}(K_{pp\sigma} - K_{pp\pi}), \notag\\
q_3 &= \tfrac{1}{4}K_{pp\sigma} + \tfrac{3}{4}K_{pp\pi}, \notag\\
q_4 &= K_{pp\sigma}.
\label{eq:q_params}
\end{align}

\begin{align}
k_0 &= \tfrac{3}{2}V_{pd\pi}\,cs^2
      - \tfrac{\sqrt{3}}{2}V_{pd\sigma}\,c\!\left(s^2 - \tfrac{1}{2}c^2\right),\notag\\
k_1 &= \tfrac{1}{2}V_{pd\sigma}\,c\!\left(s^2 - \tfrac{1}{2}c^2\right)
      - \tfrac{\sqrt{3}}{2}V_{pd\pi}\,cs^2,\notag\\
k_2 &= V_{pd\sigma}\,s\!\left(s^2 - \tfrac{1}{2}c^2\right)
      + \sqrt{3}\,V_{pd\pi}\,c^2s,\notag\\
k_3 &= V_{pd\pi}\,c,\notag\\
k_4 &= -\tfrac{3\sqrt{3}}{8}V_{pd\sigma}\,c^3
      - \tfrac{1}{2}V_{pd\pi}\,c\!\left(1-\tfrac{3}{2}c^2\right),\notag\\
k_5 &= -\tfrac{3}{8}V_{pd\sigma}\,c^3
      - \tfrac{\sqrt{3}}{2}V_{pd\pi}\,c\!\left(1-\tfrac{1}{2}c^2\right),\notag\\
k_6 &= \tfrac{3}{4}V_{pd\sigma}\,c^2s
      - \tfrac{\sqrt{3}}{2}V_{pd\pi}\,c^2s,\notag\\
k_7 &= \tfrac{3}{8}V_{pd\sigma}\,c^3
      + \tfrac{\sqrt{3}}{2}V_{pd\pi}\,c\!\left(1-\tfrac{1}{2}c^2\right),\notag\\
k_8 &= -\tfrac{\sqrt{3}}{2}V_{pd\sigma}\,c^3 - V_{pd\pi}\,cs^2,\notag\\
k_9 &= -\tfrac{\sqrt{3}}{8}V_{pd\sigma}\,c^3
      + \tfrac{1}{2}V_{pd\pi}\,c\!\left(1+\tfrac{1}{2}c^2\right),\notag\\
k_{10} &= -\tfrac{\sqrt{3}}{4}V_{pd\sigma}\,c^2s
         + \tfrac{1}{2}V_{pd\pi}\,c^2s,\notag\\
k_{11} &= \sqrt{3}\,V_{pd\sigma}\,c^2s
         + V_{pd\pi}\,(1-2c^2)\,s,\notag\\
k_{12} &= \tfrac{\sqrt{3}}{4}V_{pd\sigma}\,c^2s
         + V_{pd\pi}\!\left(1-\tfrac{1}{2}c^2\right)s,\notag\\
k_{13} &= -\tfrac{\sqrt{3}}{2}V_{pd\sigma}\,cs^2
         - \tfrac{1}{2}V_{pd\pi}\,(1-2s^2)\,c,\notag\\
k_{14} &= V_{pd\pi}\,s,\notag\\
k_{15} &= \tfrac{3\sqrt{3}}{4}V_{pd\sigma}\,c^2s
         + V_{pd\pi}\!\left(1-\tfrac{3}{2}c^2\right)s,\notag\\
k_{16} &= -\tfrac{3}{2}V_{pd\sigma}\,cs^2
         - \tfrac{\sqrt{3}}{2}V_{pd\pi}\,(1-2s^2)\,c.
\label{eq:k_params}
\end{align}

\begin{align}
t_0  &=  \tfrac{1}{4}V_{dd\sigma} + \tfrac{3}{4}V_{dd\delta},\notag\\
t_1  &= -\tfrac{3}{8}V_{dd\sigma} + \tfrac{3}{8}V_{dd\delta},\notag\\
t_2  &=  \tfrac{\sqrt{3}}{8}(V_{dd\sigma} - V_{dd\delta}),\notag\\
t_3  &=  V_{dd\pi},\notag\\
t_4  &=  \tfrac{9}{16}V_{dd\sigma} + \tfrac{1}{4}V_{dd\pi}
        + \tfrac{3}{16}V_{dd\delta},\notag\\
t_5  &= -\tfrac{3\sqrt{3}}{16}V_{dd\sigma} + \tfrac{\sqrt{3}}{4}V_{dd\pi}
        - \tfrac{\sqrt{3}}{16}V_{dd\delta},\notag\\
t_6  &=  \tfrac{3}{4}V_{dd\sigma} + \tfrac{1}{4}V_{dd\delta},\notag\\
t_7  &=  \tfrac{3}{16}V_{dd\sigma} + \tfrac{3}{4}V_{dd\pi}
        + \tfrac{1}{16}V_{dd\delta},\notag\\
t_8  &=  V_{dd\delta},\notag\\
t_9  &=  \tfrac{3}{4}V_{dd\pi} + \tfrac{1}{4}V_{dd\delta},\notag\\
t_{10} &= \tfrac{\sqrt{3}}{4}(V_{dd\pi} - V_{dd\delta}),\notag\\
t_{11} &=  t_{3},\notag\\
t_{12} &=  \tfrac{1}{4}V_{dd\pi} + \tfrac{3}{4}V_{dd\delta}.
\label{eq:t_params}
\end{align}

\begin{align}
u_0  &= \tfrac{1}{4}K_{dd\sigma} + \tfrac{3}{4}K_{dd\delta},\notag\\
u_1  &= -\tfrac{3}{8}K_{dd\sigma} + \tfrac{3}{8}K_{dd\delta},\notag\\
u_2  &= -\tfrac{\sqrt{3}}{8}K_{dd\sigma}
        + \tfrac{\sqrt{3}}{8}K_{dd\delta},\notag\\
u_3  &=  \tfrac{9}{16}K_{dd\sigma} + \tfrac{1}{4}K_{dd\pi}
        + \tfrac{3}{16}K_{dd\delta},\notag\\
u_4  &=  K_{dd\pi},\notag\\
u_5  &=  \tfrac{3\sqrt{3}}{16}K_{dd\sigma}
        - \tfrac{\sqrt{3}}{4}K_{dd\pi}
        + \tfrac{\sqrt{3}}{16}K_{dd\delta},\notag\\
u_6  &=  \tfrac{3}{16}K_{dd\sigma} + \tfrac{3}{4}K_{dd\pi}
        + \tfrac{1}{16}K_{dd\delta},\notag\\
u_7  &=  \tfrac{3}{4}K_{dd\sigma} + \tfrac{1}{4}K_{dd\delta},\notag\\
u_8  &=  \tfrac{1}{4}K_{dd\pi} + \tfrac{3}{4}K_{dd\delta},\notag\\
u_9  &= \tfrac{\sqrt{3}}{4}(K_{dd\pi} - K_{dd\delta}),\notag\\
u_{10} &= \tfrac{3}{4}K_{dd\pi} + \tfrac{1}{4}K_{dd\delta},\notag\\
u_{11} &=  K_{dd\delta}.
\label{eq:u_params}
\end{align}

The inter-sublayer X--X hopping amplitudes $r_i$ are obtained from the same $V_{pp\sigma}$ and $V_{pp\pi}$ SK integrals evaluated along the inter-sublayer bond direction $\vec{l_i}$. Defining the geometric factor $d=\cos^2\theta+4\sin^2\theta$, the eight independent amplitudes are
\begin{align}
r_0 &= \tfrac{3c^2}{4d}\,V_{pp\sigma}
      + \!\left(1 - \tfrac{3c^2}{4d}\right)V_{pp\pi},\notag\\
r_1 &= V_{pp\pi},\notag\\
r_2 &= \tfrac{\sqrt{3}\,c^2}{4d}\!\left(V_{pp\sigma} - V_{pp\pi}\right),\notag\\
r_3 &= -\tfrac{\sqrt{3}\,cs}{d}\!\left(V_{pp\sigma} - V_{pp\pi}\right),\notag\\
r_4 &= \tfrac{c^2}{4d}\,V_{pp\sigma}
      + \!\left(1 - \tfrac{c^2}{4d}\right)V_{pp\pi},\notag\\
r_5 &= \tfrac{c^2}{d}\,V_{pp\sigma}
      + \!\left(1 - \tfrac{c^2}{d}\right)V_{pp\pi},\notag\\
r_6 &= -\tfrac{cs}{d}\!\left(V_{pp\sigma} - V_{pp\pi}\right),\notag\\
r_7 &= \tfrac{4s^2}{d}\,V_{pp\sigma}
      + \!\left(1 - \tfrac{4s^2}{d}\right)V_{pp\pi}.
\label{eq:r_params}
\end{align}

%------------------------------------------------------------
\bibliographystyle{unsrt}
\bibliography{manu}

@article{Wei2023,
author={Wei, Xingbin
and Yang, Lu
and Bao, Jinlin},
title={Effect of Non-Metallic Doping and Tensile Strain on Photoelectric Properties of 1T-ZrS2 Monolayer},
journal={Russian Journal of Physical Chemistry A},
year={2023},
month={Nov},
day={01},
volume={97},
number={11},
pages={2501-2509},
issn={1531-863X},
doi={10.1134/S0036024423110353},
url={https://doi.org/10.1134/S0036024423110353}
}

@article{Mattinen2019,
    author = {Mattinen, Miika and Popov, Georgi and Vehkam\"aki, Marko and King, Peter J. and Mizohata, Kenichiro and Jalkanen, Pasi and R\"ais\"anen, Jyrki and Leskel\"a, Markku and Ritala, Mikko},
    title = {Atomic Layer Deposition of Emerging 2D Semiconductors, HfS2 and ZrS2, for Optoelectronics},
    journal = {Chemistry of Materials},
    volume = {31},
    number = {15},
    pages = {5713-5724},
    year = {2019},
    month = {08},
    issn = {0897-4756},
    doi = {10.1021/acs.chemmater.9b01688},
    url = {https://doi.org/10.1021/acs.chemmater.9b01688}
}

@article{vasp93,
  title = {Ab initio molecular dynamics for liquid metals},
  author = {Kresse, G. and Hafner, J.},
  journal = {Phys. Rev. B},
  volume = {47},
  issue = {1},
  pages = {558--561},
  numpages = {0},
  year = {1993},
  publisher = {American Physical Society},
  doi = {10.1103/PhysRevB.47.558}
}

@article{vasp96,
  title = {Efficient iterative schemes for ab initio total-energy calculations using a plane-wave basis set},
  author = {Kresse, G. and Furthm\"uller, J.},
  journal = {Phys. Rev. B},
  volume = {54},
  issue = {16},
  pages = {11169--11186},
  numpages = {0},
  year = {1996},
  publisher = {American Physical Society},
  doi = {10.1103/PhysRevB.54.11169}
}

@article{PBE,
  title={Generalized gradient approximation made simple},
  author={Perdew, John P and Burke, Kieron and Ernzerhof, Matthias},
  journal={Physical review letters},
  volume={77},
  number={18},
  pages={3865},
  year={1996},
  publisher={APS}
}

@article{Cappelluti2013,
  title = {Tight-binding model and direct-gap/indirect-gap transition in single-layer and multilayer MoS${}_{2}$},
  author = {Cappelluti, E. and Rold\'an, R. and Silva-Guill\'en, J. A. and Ordej\'on, P. and Guinea, F.},
  journal = {Phys. Rev. B},
  volume = {88},
  issue = {7},
  pages = {075409},
  numpages = {18},
  year = {2013},
  publisher = {American Physical Society},
  doi = {10.1103/PhysRevB.88.075409}
}

@article{Slater-Koster,
  title = {Simplified LCAO Method for the Periodic Potential Problem},
  author = {Slater, J. C. and Koster, G. F.},
  journal = {Phys. Rev.},
  volume = {94},
  issue = {6},
  pages = {1498--1524},
  numpages = {0},
  year = {1954},
  publisher = {American Physical Society},
  doi = {10.1103/PhysRev.94.1498}
}

@article{Peng2024,
  title = {Modified tight-binding model for strain effects in monolayer transition metal dichalcogenides},
  author = {Peng, Zhiwei and Guan, Zhizi and Wang, Hongfei and Srolovitz, David J. and Lei, Dangyuan},
  journal = {Phys. Rev. B},
  volume = {109},
  issue = {24},
  pages = {245412},
  numpages = {13},
  year = {2024},
  month = {Jun},
  publisher = {American Physical Society},
  doi = {10.1103/PhysRevB.109.245412},
  url = {https://link.aps.org/doi/10.1103/PhysRevB.109.245412}
}

@article{Roldan2014,
doi = {10.1088/2053-1583/1/3/034003},
url = {https://dx.doi.org/10.1088/2053-1583/1/3/034003},
year = {2014},
month = {nov},
publisher = {IOP Publishing},
volume = {1},
number = {3},
pages = {034003},
author = {Roldán, R and López-Sancho, M P and Guinea, F and Cappelluti, E and Silva-Guillén, J A and Ordejón, P},
title = {Momentum dependence of spin–orbit interaction effects in single-layer and multi-layer transition metal dichalcogenides},
journal = {2D Materials}
}

@article{Dias2018,
  title = {Band structure of monolayer transition-metal dichalcogenides and topological properties of their nanoribbons: Next-nearest-neighbor hopping},
  author = {Dias, A. C. and Qu, Fanyao and Azevedo, David L. and Fu, Jiyong},
  journal = {Phys. Rev. B},
  volume = {98},
  issue = {7},
  pages = {075202},
  numpages = {19},
  year = {2018},
  month = {Aug},
  publisher = {American Physical Society},
  doi = {10.1103/PhysRevB.98.075202},
  url = {https://link.aps.org/doi/10.1103/PhysRevB.98.075202}
}

@article{PRB.100.165304,
  title = {Electronic and hyperbolic dielectric properties of $\mathrm{Zr}{\mathrm{S}}_{2}/\mathrm{Hf}{\mathrm{S}}_{2}$ heterostructures},
  author = {Zhang, Liwei and Yu, Weiyang and Wang, Qin and Ou, Jun-Yu and Wang, Baoji and Tang, Gang and Jia, Xingtao and Yang, Xuefeng and Wang, Guodong and Cai, Xiaolin},
  journal = {Phys. Rev. B},
  volume = {100},
  issue = {16},
  pages = {165304},
  numpages = {10},
  year = {2019},
  month = {Oct},
  publisher = {American Physical Society},
  doi = {10.1103/PhysRevB.100.165304},
  url = {https://link.aps.org/doi/10.1103/PhysRevB.100.165304}
}

@article{PRM.3.074001,
  title = {Electronic and optical excitations of two-dimensional ${\mathrm{ZrS}}_{2}$ and ${\mathrm{HfS}}_{2}$ and their heterostructure},
  author = {Lau, Ka Wai and Cocchi, Caterina and Draxl, Claudia},
  journal = {Phys. Rev. Mater.},
  volume = {3},
  issue = {7},
  pages = {074001},
  numpages = {9},
  year = {2019},
  month = {Jul},
  publisher = {American Physical Society},
  doi = {10.1103/PhysRevMaterials.3.074001},
  url = {https://link.aps.org/doi/10.1103/PhysRevMaterials.3.074001}
}

@book{f_wmse,
  title={Numerical Recipes in C: The Art of Scientific Computing},
  author={Press, William H and Teukolsky, Saul A and Vetterling, William T and Flannery, Brian P},
  volume={2},
  year={1992},
  publisher={Cambridge University Press},
  address={Cambridge}
}

@article{murray1972,
doi = {10.1088/0022-3719/5/7/006},
url = {https://doi.org/10.1088/0022-3719/5/7/006},
year = {1972},
month = {apr},
publisher = {},
volume = {5},
number = {7},
pages = {746},
author = {R B Murray and R A Bromley and A D Yoffe},
title = {The band structures of some transition metal dichalcogenides. II. Group IVA; octahedral coordination},
journal = {Journal of Physics C: Solid State Physics},
}

@article{Silva-Guillen2016,
AUTHOR = {Silva-Guillén, Jose Ángel and San-Jose, Pablo and Roldán, Rafael},
TITLE = {Electronic Band Structure of Transition Metal Dichalcogenides from Ab Initio and Slater–Koster Tight-Binding Model},
JOURNAL = {Applied Sciences},
VOLUME = {6},
YEAR = {2016},
NUMBER = {10},
ARTICLE-NUMBER = {284},
URL = {https://www.mdpi.com/2076-3417/6/10/284},
ISSN = {2076-3417},
DOI = {10.3390/app6100284}
}

@article{Ghosh2025,
  title = {Choosing tight-binding models for accurate optoelectronic responses},
  author = {Ghosh, Andreas and Schankler, Aaron M. and Rappe, Andrew M.},
  journal = {Phys. Rev. B},
  volume = {111},
  issue = {12},
  pages = {125203},
  numpages = {8},
  year = {2025},
  month = {Mar},
  publisher = {American Physical Society},
  doi = {10.1103/PhysRevB.111.125203},
  url = {https://link.aps.org/doi/10.1103/PhysRevB.111.125203}
}

@article{Lado_2016,
doi = {10.1088/2053-1583/3/3/035023},
url = {https://doi.org/10.1088/2053-1583/3/3/035023},
year = {2016},
month = {sep},
publisher = {IOP Publishing},
volume = {3},
number = {3},
pages = {035023},
author = {Lado, J L and Fernández-Rossier, J},
title = {Landau levels in 2D materials using Wannier Hamiltonians obtained by first principles},
journal = {2D Materials}
}

@article{kubo1957,
  title={Statistical-mechanical theory of irreversible processes. II. Response to thermal disturbance},
  author={Kubo, Ryogo and Yokota, Mario and Nakajima, Sadao},
  journal={Journal of the Physical Society of Japan},
  volume={12},
  number={11},
  pages={1203--1211},
  year={1957},
  publisher={The Physical Society of Japan}
}

@article{Greenwood_1958,
doi = {10.1088/0370-1328/71/4/306},
url = {https://doi.org/10.1088/0370-1328/71/4/306},
year = {1958},
month = {apr},
publisher = {},
volume = {71},
number = {4},
pages = {585},
author = {D A Greenwood},
title = {The Boltzmann Equation in the Theory of Electrical Conduction in Metals},
journal = {Proceedings of the Physical Society}
}

@article{Torma2023,
  title = {Conductivity in flat bands from the Kubo-Greenwood formula},
  author = {Huhtinen, Kukka-Emilia and T\"orm\"a, P\"aivi},
  journal = {Phys. Rev. B},
  volume = {108},
  issue = {15},
  pages = {155108},
  numpages = {9},
  year = {2023},
  month = {Oct},
  publisher = {American Physical Society},
  doi = {10.1103/PhysRevB.108.155108},
  url = {https://link.aps.org/doi/10.1103/PhysRevB.108.155108}
}

@article{Resta2011,
author={Resta, R.},
title={The insulating state of matter: a geometrical theory},
journal={The European Physical Journal B},
year={2011},
month={Jan},
day={01},
volume={79},
number={2},
pages={121-137},
issn={1434-6036},
doi={10.1140/epjb/e2010-10874-4},
url={https://doi.org/10.1140/epjb/e2010-10874-4}
}

@article{Xiao2010,
  title = {Berry phase effects on electronic properties},
  author = {Xiao, Di and Chang, Ming-Che and Niu, Qian},
  journal = {Rev. Mod. Phys.},
  volume = {82},
  issue = {3},
  pages = {1959--2007},
  numpages = {0},
  year = {2010},
  month = {Jul},
  publisher = {American Physical Society},
  doi = {10.1103/RevModPhys.82.1959},
  url = {https://link.aps.org/doi/10.1103/RevModPhys.82.1959}
}

@article{Provost1980,
  author  = {Provost, J. P. and Vallee, G.},
  title   = {Riemannian structure on manifolds of quantum states},
  journal = {Commun. Math. Phys.},
  volume  = {76},
  pages   = {289},
  year    = {1980},
  doi     = {10.1007/BF02193559}
}

@article{Ozawa2018,
  author  = {Ozawa, T. and Goldman, N.},
  title   = {Extracting the quantum metric tensor through periodic driving},
  journal = {Phys. Rev. B},
  volume  = {97},
  pages   = {201117(R)},
  year    = {2018},
  doi     = {10.1103/PhysRevB.97.201117}
}

@article{Souza2000,
  author  = {Souza, I. and Wilkens, T. and Martin, R. M.},
  title   = {Polarization and localization in insulators: generating function approach},
  journal = {Phys. Rev. B},
  volume  = {62},
  pages   = {1666},
  year    = {2000},
  doi     = {10.1103/PhysRevB.62.1666}
}

@article{Verma2025,
  title = {Framework to Measure Quantum Metric from Step Response},
  author = {Verma, Nishchhal and Queiroz, Raquel},
  journal = {Phys. Rev. Lett.},
  volume = {134},
  issue = {10},
  pages = {106403},
  numpages = {6},
  year = {2025},
  month = {Mar},
  publisher = {American Physical Society},
  doi = {10.1103/PhysRevLett.134.106403},
  url = {https://link.aps.org/doi/10.1103/PhysRevLett.134.106403}
}

@article{Yu2025,
author={Yu, Jiabin
and Bernevig, B. Andrei
and Queiroz, Raquel
and Rossi, Enrico
and T{\"o}rm{\"a}, P{\"a}ivi
and Yang, Bohm-Jung},
title={Quantum geometry in quantum materials},
journal={npj Quantum Materials},
year={2025},
month={Oct},
day={10},
volume={10},
number={1},
pages={101},
doi={10.1038/s41535-025-00801-3},
url={https://doi.org/10.1038/s41535-025-00801-3}
}

@article{Ghosh2024,
author = {Barun Ghosh  and Yugo Onishi  and Su-Yang Xu  and Hsin Lin  and Liang Fu  and Arun Bansil },
title = {Probing quantum geometry through optical conductivity and magnetic circular dichroism},
journal = {Science Advances},
volume = {10},
number = {51},
year = {2024},
doi = {10.1126/sciadv.ado1761},
URL = {https://www.science.org/doi/abs/10.1126/sciadv.ado1761}
}

@article{Ridolfi2015,
doi = {10.1088/0953-8984/27/36/365501},
url = {https://doi.org/10.1088/0953-8984/27/36/365501},
year = {2015},
month = {aug},
publisher = {IOP Publishing},
volume = {27},
number = {36},
pages = {365501},
author = {Ridolfi, E and Le, D and Rahman, T S and Mucciolo, E R and Lewenkopf, C H},
title = {A tight-binding model for MoS2 monolayers},
journal = {Journal of Physics: Condensed Matter}
}

@article{Yu2024,
author={Yu, Jiabin
and Ciccarino, Christopher J.
and Bianco, Raffaello
and Errea, Ion
and Narang, Prineha
and Bernevig, B. Andrei},
title={Non-trivial quantum geometry and the strength of electron--phonon coupling},
journal={Nature Physics},
year={2024},
month={Aug},
day={01},
volume={20},
number={8},
pages={1262-1268},
issn={1745-2481},
doi={10.1038/s41567-024-02486-0},
url={https://doi.org/10.1038/s41567-024-02486-0}
}

@article{Torma2023essay,
  title = {Essay: Where Can Quantum Geometry Lead Us?},
  author = {T\"orm\"a, P\"aivi},
  journal = {Phys. Rev. Lett.},
  volume = {131},
  issue = {24},
  pages = {240001},
  numpages = {7},
  year = {2023},
  month = {Dec},
  publisher = {American Physical Society},
  doi = {10.1103/PhysRevLett.131.240001},
  url = {https://link.aps.org/doi/10.1103/PhysRevLett.131.240001}
}

@article{Ahn2022,
author={Ahn, Junyeong
and Guo, Guang-Yu
and Nagaosa, Naoto
and Vishwanath, Ashvin},
title={Riemannian geometry of resonant optical responses},
journal={Nature Physics},
year={2022},
month={Mar},
day={01},
volume={18},
number={3},
pages={290-295},
issn={1745-2481},
doi={10.1038/s41567-021-01465-z},
url={https://doi.org/10.1038/s41567-021-01465-z}
}

@article{Nagaosa2010,
  title = {Anomalous Hall effect},
  author = {Nagaosa, Naoto and Sinova, Jairo and Onoda, Shigeki and MacDonald, A. H. and Ong, N. P.},
  journal = {Rev. Mod. Phys.},
  volume = {82},
  issue = {2},
  pages = {1539--1592},
  numpages = {0},
  year = {2010},
  month = {May},
  publisher = {American Physical Society},
  doi = {10.1103/RevModPhys.82.1539},
  url = {https://link.aps.org/doi/10.1103/RevModPhys.82.1539}
}

@article{Ezawa2024,
  title = {Analytic approach to quantum metric and optical conductivity in Dirac models with parabolic mass in arbitrary dimensions},
  author = {Ezawa, Motohiko},
  journal = {Phys. Rev. B},
  volume = {110},
  issue = {19},
  pages = {195437},
  numpages = {11},
  year = {2024},
  month = {Nov},
  publisher = {American Physical Society},
  doi = {10.1103/PhysRevB.110.195437},
  url = {https://link.aps.org/doi/10.1103/PhysRevB.110.195437}
}

@article{
Nishchhal2024,
author = {Nishchhal Verma  and Raquel Queiroz },
title = {Instantaneous response and quantum geometry of insulators},
journal = {Proceedings of the National Academy of Sciences},
volume = {122},
number = {49},
pages = {e2405837122},
year = {2025},
doi = {10.1073/pnas.2405837122},
URL = {https://www.pnas.org/doi/abs/10.1073/pnas.2405837122}}

@article{
Kim2025,
author = {Sunje Kim  and Yoonah Chung  and Yuting Qian  and Soobin Park  and Chris Jozwiak  and Eli Rotenberg  and Aaron Bostwick  and Keun Su Kim  and Bohm-Jung Yang },
title = {Direct measurement of the quantum metric tensor in solids},
journal = {Science},
volume = {388},
number = {6751},
pages = {1050-1054},
year = {2025},
doi = {10.1126/science.ado6049},
URL = {https://www.science.org/doi/abs/10.1126/science.ado6049}}

@article{Simon2020,
  title = {Contrasting lattice geometry dependent versus independent quantities: Ramifications for Berry curvature, energy gaps, and dynamics},
  author = {Simon, Steven H. and Rudner, Mark S.},
  journal = {Phys. Rev. B},
  volume = {102},
  issue = {16},
  pages = {165148},
  numpages = {13},
  year = {2020},
  month = {Oct},
  publisher = {American Physical Society},
  doi = {10.1103/PhysRevB.102.165148},
  url = {https://link.aps.org/doi/10.1103/PhysRevB.102.165148}
}

@misc{Telle2026,
      title={Optimally embedded tight binding for reproducing geometry dependent observables}, 
      author={Jonas J Telle and Gunnar F Lange},
      year={2026},
      eprint={2608.07684},
      archivePrefix={arXiv},
      primaryClass={cond-mat.mes-hall},
      url={https://arxiv.org/abs/2608.07684}, 
}

@article{
Gao2023,
author = {Anyuan Gao  and Yu-Fei Liu  and Jian-Xiang Qiu  and Barun Ghosh  and Thaís V. Trevisan  and Yugo Onishi  and Chaowei Hu  and Tiema Qian  and Hung-Ju Tien  and Shao-Wen Chen  and Mengqi Huang  and Damien Bérubé  and Houchen Li  and Christian Tzschaschel  and Thao Dinh  and Zhe Sun  and Sheng-Chin Ho  and Shang-Wei Lien  and Bahadur Singh  and Kenji Watanabe  and Takashi Taniguchi  and David C. Bell  and Hsin Lin  and Tay-Rong Chang  and Chunhui Rita Du  and Arun Bansil  and Liang Fu  and Ni Ni  and Peter P. Orth  and Qiong Ma  and Su-Yang Xu },
title = {Quantum metric nonlinear Hall effect in a topological antiferromagnetic heterostructure},
journal = {Science},
volume = {381},
number = {6654},
pages = {181-186},
year = {2023},
doi = {10.1126/science.adf1506},
URL = {https://www.science.org/doi/abs/10.1126/science.adf1506}}

@article{Raquel2026,
author={Verma, Nishchhal
and Moll, Philip J. W.
and Holder, Tobias
and Queiroz, Raquel},
title={Quantum geometry and the hidden scales in materials},
journal={Nature Reviews Physics},
year={2026},
month={Apr},
day={01},
volume={8},
number={4},
pages={226-239},
doi={10.1038/s42254-026-00923-y},
url={https://doi.org/10.1038/s42254-026-00923-y}
}

@article{Wang2018,
    author = {Wang, Denggui and Meng, Junhua and Zhang, Xingwang and Guo, Gencai and Yin, Zhigang and Liu, Heng and Cheng, Likun and Gao, Menglei and You, Jingbi and Wang, Ruzhi},
    title = {Selective Direct Growth of Atomic Layered HfS2 on Hexagonal Boron Nitride for High Performance Photodetectors},
    journal = {Chemistry of Materials},
    volume = {30},
    number = {11},
    pages = {3819-3826},
    year = {2018},
    month = {05},
    issn = {0897-4756},
    doi = {10.1021/acs.chemmater.8b01091},
    url = {https://doi.org/10.1021/acs.chemmater.8b01091},
}

@article{Tanthirige2019,
title={Intrinsic Photoconductivity of Few-layered ZrS2 Phototransistors via Multiterminal Measurements}, 
volume={1}, 
url={https://journals.bilpubgroup.com/index.php/ssid/article/view/1526},
DOI={10.30564/ssid.v1i2.1526},
number={2},
journal={Semiconductor Science and Information Devices},
author={Tanthirige, Rukshan M. and Garcia, Carlos and Ghosh, Saikat and II, Frederick Jackson and Nash, Jawnaye and Rosenmann, Daniel and Divan, Ralu and Stan, Liliana and Sumant, Anirudha V. and McGill, Stephen A. and Ray, Paresh C. and Pradhan, Nihar R.},
year={2019},
month={Oct.},
pages={19–28} 
}

@article{Orujlu2026,
title = {Exploring the layer-dependent characteristics of ZrS2 for flexible optoelectronic and photovoltaic devices},
journal = {Solid State Communications},
volume = {412},
pages = {116412},
year = {2026},
issn = {0038-1098},
doi = {https://doi.org/10.1016/j.ssc.2026.116412},
author = {N.A. Orujlu and N.A. Ismayilova and H. Ozisik and E. Deligoz and Y.I. Aliyev and S.H. Jabarov}
}

@article{Zhang2022,
title = {Effects of the in-plane uniaxial and biaxial strains on the structural and electronic properties of the monolayer ZrS2: A first-principles investigation},
journal = {Thin Solid Films},
volume = {755},
pages = {139343},
year = {2022},
issn = {0040-6090},
doi = {https://doi.org/10.1016/j.tsf.2022.139343},
author = {Yan Zhang and Li Duan and Ji-Bin Fan and Lei Ni},
}

@article{Liu2025,
  title     = "Quantum geometry in condensed matter",
  author    = "Liu, Tianyu and Qiang, Xiao-Bin and Lu, Hai-Zhou and Xie, X C",
  journal   = "Natl. Sci. Rev.",
  publisher = "Oxford University Press (OUP)",
  volume    =  12,
  number    =  3,
  month     =  mar,
  year      =  2025,
}

@misc{Gao2025,
      title={Quantum Geometry Phenomena in Condensed Matter Systems}, 
      author={Anyuan Gao and Naoto Nagaosa and Ni Ni and Su-Yang Xu},
      year={2025},
      eprint={2508.00469},
      archivePrefix={arXiv},
      primaryClass={cond-mat.str-el},
      url={https://arxiv.org/abs/2508.00469}, 
}

@article{Marzari2012,
  title = {Maximally localized Wannier functions: Theory and applications},
  author = {Marzari, Nicola and Mostofi, Arash A. and Yates, Jonathan R. and Souza, Ivo and Vanderbilt, David},
  journal = {Rev. Mod. Phys.},
  volume = {84},
  issue = {4},
  pages = {1419--1475},
  numpages = {0},
  year = {2012},
  month = {Oct},
  publisher = {American Physical Society},
  doi = {10.1103/RevModPhys.84.1419},
  url = {https://link.aps.org/doi/10.1103/RevModPhys.84.1419}
}

@article{Papaconstantopoulos2003,
doi = {10.1088/0953-8984/15/10/201},
url = {https://doi.org/10.1088/0953-8984/15/10/201},
year = {2003},
month = {mar},
publisher = {},
volume = {15},
number = {10},
pages = {R413},
author = {D A Papaconstantopoulos and M J Mehl},
title = {The Slater–Koster tight-binding method: a computationally efficient and accurate
approach},
journal = {Journal of Physics: Condensed Matter},
}

@article{Jorissen2024,
	title = {Comparative analysis of tight-binding models for transition metal dichalcogenides},
	pages = {004},
	author = {Jorissen, Bert and Covaci, Lucian and Partoens, Bart},
	journal = {SciPost Phys. Core},
	volume = {7},
	year = {2024},
	publisher = {SciPost},
	doi = {10.21468/SciPostPhysCore.7.1.004},
	url = {https://scipost.org/10.21468/SciPostPhysCore.7.1.004}
}

\end{document}